\documentclass[fleqn,usenatbib]{mnras}

\usepackage{newtxtext,newtxmath}

\usepackage[T1]{fontenc}
\usepackage{ae,aecompl}

\usepackage{graphicx}	
\usepackage{amsmath}	
\usepackage{multicol} 
\usepackage{multirow}
\usepackage{comment}
\usepackage{array}
\usepackage{soul}
\hypersetup{colorlinks=true,citecolor=blue,linkcolor=purple,filecolor=black,runcolor=black,breaklinks=true}
\usepackage{etoolbox}
\usepackage{ae,aecompl}
\usepackage[usenames]{color}
\usepackage{hyperref}
\usepackage{threeparttable}
\usepackage{mathrsfs}
\usepackage{scalerel}
\usepackage{longtable,booktabs,threeparttablex}

\definecolor{purple}{RGB}{160,32,240}

\definecolor{red}{RGB}{225,50,50}

\newcommand{\HST}{\emph{HST}}
\newcommand{\JWST}{\emph{JWST}}
\newcommand{\Spitzer}{\emph{Spitzer}}
\newcommand{\Muv}{\ensuremath{\mathrm{M}_{\mathrm{UV}}^{ }}}
\newcommand{\muv}{\ensuremath{m_{\mathrm{UV}}^{ }}}

\newcommand{\logten}{\ensuremath{\log_{10}}}

\newcommand{\Lya}{\ensuremath{\mathrm{Ly}\alpha}}
\newcommand{\zLya}{\ensuremath{\mathrm{z}_{_{\mathrm{Ly\alpha}}}}}

\newcommand{\xHI}{\ensuremath{\mathrm{x}_{\mathrm{HI}}}}

\newcommand{\xiion}{\ensuremath{\xi_{\mathrm{ion}}^{\ast}}}
\newcommand{\logxiion}{\ensuremath{\log_{10}[\xi_{\mathrm{ion}}^{\ast} / (\mathrm{erg}^{-1}\ \mathrm{Hz})]}}

\newcommand{\fesc}{\ensuremath{f_{\mathrm{esc}}}}
\newcommand{\OIIIHb}{[OIII]+H\ensuremath{\beta}}

\newcommand{\muEW}{\ensuremath{\mu_{\mathrm{EW}}}}
\newcommand{\sigmaEW}{\ensuremath{\sigma_{\mathrm{EW}}}}

\newcommand{\Msol}{\ensuremath{\mathrm{M}_{\odot}}}
\newcommand{\Mstar}{\ensuremath{\mathrm{M}_{\ast}}}
\newcommand{\logMstar}{\ensuremath{\log_{10}\left(\mathrm{M}_{\ast}/\mathrm{M}_{\odot}\right)}}

\newcolumntype{P}[1]{>{\centering\arraybackslash}p{#1}}
\newcommand\Tstrut{\rule{0pt}{2.6ex}}         
\newcommand\Bstrut{\rule[-1.2ex]{0pt}{0pt}}   

\title[MMT Spectroscopy of Lyman-alpha at z$\simeq$7]{MMT Spectroscopy of Lyman-alpha at z$\simeq$7: Evidence for Accelerated Reionization Around Massive Galaxies}

\author[R. Endsley et al.]{
Ryan Endsley$^{1}$\thanks{E-mail: rendsley@email.arizona.edu},
Daniel P. Stark$^{1}$, 
St\'ephane Charlot$^{2}$, 
Jacopo Chevallard$^{2}$,
\newauthor{
Brant Robertson$^{3}$,
Rychard J. Bouwens$^{4}$,
Mauro Stefanon$^{4}$}
\\
$^{1}$Steward Observatory, University of Arizona, 933 N Cherry Ave, Tucson, AZ 85721 USA\\
$^{2}$Sorbonne Universit\'es, UPMC-CNRS, UMR7095, Institut d'Astrophysique de Paris, F-75014, Paris, France\\
$^{3}$Department of Astronomy and Astrophysics, University of California, Santa Cruz, 1156 High Street, Santa Cruz, CA 95064 USA\\
$^{4}$Leiden Observatory, Leiden University, NL-2300 RA Leiden, The Netherlands
}

\date{Accepted XXX. Received YYY; in original form ZZZ}

\pubyear{2020}

\begin{document}
\label{firstpage}
\pagerange{\pageref{firstpage}--\pageref{lastpage}}
\maketitle


\begin{abstract}
Reionization-era galaxies tend to exhibit weak Ly$\alpha$ emission, 
likely reflecting attenuation from an increasingly neutral IGM. 
Recent observations have begun to reveal exceptions to this picture, 
with strong \Lya{} emission now known in four of the 
most massive z=7--9 galaxies in the CANDELS fields, all of which also exhibit intense \OIIIHb{} emission (EW$>$800 \AA{}). 
To better understand why \Lya{} is anomalously strong in a subset 
of massive $z\simeq 7-9$ galaxies, we have initiated an MMT/Binospec survey targeting a larger sample (N=22) of similarly luminous ($\simeq$1--6 L$^{\ast}_{\mathrm{UV}}$) z$\simeq$7 galaxies selected over very wide-area fields ($\sim$3 deg$^2$).
We confidently ($>$7$\sigma$) detect \Lya{} in 78\% (7/9) of galaxies with strong \OIIIHb{} emission (EW$>$800 \AA{}) as opposed to only 8\% (1/12) of galaxies with more moderate (EW=200-800 \AA{}) \OIIIHb{}.
We argue that the higher \Lya{} EWs of the strong \OIIIHb{} population likely reflect enhanced ionizing photon production efficiency owing to their large sSFRs ($\gtrsim$30 Gyr$^{-1}$).
We also find evidence that \Lya{} transmission from massive galaxies declines less rapidly over $6<z<7$ than in low-mass lensed systems.
In particular, our data suggest no strong evolution in \Lya{} transmission, 
consistent with a picture wherein massive z$\simeq$7 galaxies 
often reside in large ionized regions.  We detect three closely-separated ($R$ = 1.7 physical Mpc) z$\simeq$7 \Lya{} emitters in our sample, conceivably tracing a large ionized structure that is 
consistent with this picture.
We detect tentative evidence for an overdensity in this region, implying a large ionizing photon budget in the surrounding volume.
\end{abstract}

\begin{keywords}
galaxies: high-redshift -- cosmology: dark ages, reionization, first stars -- galaxies: evolution \end{keywords}



\defcitealias{Endsley2021}{E21}
\defcitealias{RobertsBorsani2016}{RB16}
\defcitealias{Schenker2014}{S14}

\section{Introduction}

Within the past decade, a number of concerted efforts have aimed at better understanding how and when the process of hydrogen reionization occurred.
Such information provides insight into not only the growth of structure in the very early Universe, but also the nature of ionizing sources responsible for reionization \citep[e.g.][]{LoebBarkana2001,Fan2006,Robertson2015,Bouwens2015_reionization,Finkelstein2016,Stark2016_ARAA,Dayal2018,Naidu2020}. 
Observationally, there has been great progress in identifying star-forming galaxies at z$\sim$6--10 (e.g. \citealt{Ellis2013,McLure2013,Bowler2014,Bowler2020,Atek2015b,Bouwens2015_LF,Bouwens2019,Finkelstein2015_LF,Livermore2017,Oesch2018_z10LF,Ono2018,Stefanon2017_Brightestz89,Stefanon2019}), revealing a population capable of generating copious amounts of ionizing photons in the early Universe.

The impact of these early ionizing agents on the IGM can be tracked independently using measurements of the timeline of the reionization 
process. Prominent detections of the \Lya{} and Ly$\beta$ forests in quasar spectra at z$\lesssim$6 provide a model-independent constraint suggesting that reionization was largely complete by z=5.9 (neutral fraction \xHI{}$\lesssim$10\%; \citealt{McGreer2015}).
However, the presence of a long neutral patch identified at z=5.5 \citep{Becker2015} suggests that reionization may have ended as late as z$\simeq$5.2--5.3 \citep{Kulkarni2019_reionization,Keating2020,Nasir2020}.
At yet higher redshifts, strong damping wing signatures in the earliest known quasars (z$\simeq$7.0--7.5) provide evidence for significant neutral 
hydrogen fractions (\xHI{}$\sim$40--60\%; \citealt{Greig2017,Banados2018,Davies2018,Wang2020}).
Measurements of the Thomson scattering optical depth from the CMB are consistent with this reionization timeline, suggesting a reionization midpoint of z$\sim$7.7 \citep{Planck2016_Reionization,Planck2020}.

High-redshift (z$\gtrsim$6) galaxies have also long enabled complementary constraints on the timeline of reionization owing to the resonant nature of the \Lya{} emission line \citep[e.g.][]{MiraldaEscude1998,Malhotra2004,Santos2004,Mesinger2008,Dijkstra2014}.
A number of studies have uncovered a marked decline in the fraction of star-forming galaxies showing strong \Lya{} (rest-frame equivalent width EW$>$25 \AA{}) at z$>$6, consistent with expectations of an increasingly neutral IGM \citep[e.g.][]{Fontana2010,Stark2010,Vanzella2011,Ono2012,Treu2013,Caruana2014,Pentericci2014,Pentericci2018,Schenker2014,deBarros2017,Jung2017,Hoag2019,Mason2019a}. 
In a similar manner, the volume density of narrow-band selected \Lya{} emitters has been found to decrease rapidly from z$\sim$6 to z$\sim$7 \citep[e.g.][]{Malhotra2004,Hu2010,Ouchi2010,Kashikawa2011,Konno2014,Ota2017,Zheng2017}.
Both these downturns suggest a highly neutral universe (\xHI{}$>$0.4) at z$\simeq$7 \citep[e.g.][]{Ota2017,Zheng2017,Mason2018_IGMneutralFrac,Whitler2020}, consistent with inferences from quasar spectra.

In the last five years, attention has begun to focus on the \Lya{} properties of the most massive galaxies known in the reionization era.
Recent spectroscopic observations have revealed a 100\% \Lya{} detection rate among four of the brightest\footnote{Throughout this work, we adopt M$_{\mathrm{UV}}^{\ast}$ = $-$20.6 from the z$\simeq$7 luminosity function results of \citet{Bowler2017}.} (3--4 L$_{\mathrm{UV}}^{\ast}$) and 
most massive z=7--9 galaxies selected over the CANDELS fields (\citealt{Oesch2015,Zitrin2015,RobertsBorsani2016,Stark2017}, hereafter the \citetalias{RobertsBorsani2016} sample), in marked contrast to the $\lesssim$10--20\% \Lya{} detection rate among typical ($<$L$_{\mathrm{UV}}^{\ast}$) galaxies at z$\gtrsim$7 \citep[e.g.][]{Ono2012,Treu2013,Pentericci2014,Pentericci2018,Schenker2014}.
It is not yet clear why the \Lya{} photons emerging from the \citetalias{RobertsBorsani2016} galaxies are so readily detectable given the highly neutral state of the IGM at z$>$7.
One likely possibility is that these massive galaxies trace strong overdensities \citep{Zitrin2015} and are hence situated in large ionized regions in 
the IGM \citep[e.g.][]{Furlanetto2004,Wyithe2005,Lee2007,McQuinn2007,Weinberger2018}.
These bubbles enable \Lya{} photons to redshift further into the damping wing before encountering HI, boosting the transmission \citep[e.g.][]{Mesinger2004,Mason2020}.
On the other hand, the high-EW \OIIIHb{} emission ($>$800 \AA{}) of the \citetalias{RobertsBorsani2016} sample may suggest that their \Lya{} detections were driven more by physics internal to each of the four galaxies \citep{Stark2017}.
Strong \OIIIHb{} typically signals a recent rapid upturn or burst in star formation activity, i.e. high specific star formation rate (sSFR; \citealt{Tang2019,Endsley2021}, hereafter \citetalias{Endsley2021}), giving rise to very young stellar populations.
The spectra of such galaxies are therefore dominated by extremely hot stars which produce ionizing photons very efficiently with respect to the far-UV continuum. Assuming these ionizing photons 
are reprocessed into recombination lines, this should also boost the production rate of Ly$\alpha$ relative to the rest-UV continuum.

To better understand the origin of strong \Lya{} from massive reionization-era galaxies, we have initiated an MMT/Binospec campaign targeting \Lya{} in a much larger sample of bright (1--6 L$_{\mathrm{UV}}^{\ast}$) z$\simeq$7 systems.
Here, we use the first results from our spectroscopic campaign to investigate to what extent strong \OIIIHb{} (and hence large sSFRs) enhance \Lya{} detectability among bright reionization-era galaxies.
Because our sources were selected over the very wide-area COSMOS and XMM3 fields ($\approx$1.5 deg$^2$ each), we are able to assemble a much larger sample of these rare systems relative to CANDELS ($\approx$0.2 deg$^2$ total; \citealt{Grogin2011}; \citealt{Koekemoer2011}).

We furthermore test whether our results are consistent with a picture wherein bright (and therefore likely massive; e.g. \citealt{BaroneNugent2014,Song2016}) z$\simeq$7 galaxies often reside in large highly-ionized bubbles, as would be expected if they preferentially trace overdensities. 
If this is the case, we would expect the transmission of \Lya{} photons from bright galaxies to show a substantially weaker decline at z$\gtrsim$6 relative to much fainter sources.
Recent observations suggest a sharp (factor of $\sim$10) decline in \Lya{} transmission for faint ($\sim$0.1 L$_{\mathrm{UV}}^{\ast}$) lensed galaxies between z$\simeq$6--7 \citep{Hoag2019,Fuller2020}. 
Our goal is to test whether our bright (1--6 L$_{\mathrm{UV}}^{\ast}$) z$\simeq$7 galaxy sample exhibit a less rapid decline in \Lya{} transmission, building on 
earlier work targeting massive galaxies (e.g. \citealt{Ono2012,Furusawa2016}).
For this investigation, we complement our z$\simeq$7 targets with a collection of similarly bright z$\simeq$6 sources that were observed simultaneously with the multiplex Binospec instrument.

This paper is organized as follows. 
In \S\ref{sec:observations}, we describe our observations including our sample selection, spectroscopic results, and the inferred \OIIIHb{} EWs of each source.
We present our analysis in \S\ref{sec:analysis}, including an investigation of whether \Lya{} detectability is enhanced in bright z$\simeq$7 systems with strong \OIIIHb{}.
We then discuss what our results imply for the presence of ionized regions around massive reionization-era galaxies in \S\ref{sec:discussion}.
Our main conclusions are summarized in \S\ref{sec:summary}.

All magnitudes are quoted in the AB system \citep{OkeGunn1983} and we adopt a flat $\Lambda$CDM cosmology with h=0.7, $\Omega_\mathrm{M}$ = 0.3, and $\Omega_\mathrm{\Lambda}$ = 0.7, consistent with \textit{Planck} results \citep{Planck2020}. All distances are quoted in physical units unless otherwise stated.

\section{Observations} \label{sec:observations}

We have initiated a large spectroscopic \Lya{} survey of bright ($\simeq$1--6 L$_{\mathrm{UV}}^{\ast}$) galaxies at z$\simeq$7 with MMT/Binospec. 
Our selection criteria are described in \S\ref{sec:sample_selection}.
We then detail our spectroscopic results and the photometric properties of galaxies in our spectroscopic sample in \S\ref{sec:binospec}. 
Finally, we infer the physical properties (e.g. stellar mass and \OIIIHb{} EW) of each spectroscopic target in \S\ref{sec:BEAGLE}.

\subsection{Source Selection and Photometry} \label{sec:sample_selection}

In this work, we aim to accomplish two goals focused on targeting \Lya{} in bright ($\gtrsim$L$_{\mathrm{UV}}^{\ast}$) galaxies at z$\simeq$6--7.
First, we seek to investigate whether strong \OIIIHb{} (and hence large sSFRs) lead to enhanced \Lya{} emission among bright reionization-era galaxies.
Second, we aim to quantify the evolution in the \Lya{} EW distribution of bright galaxies between z$\simeq$6--7 to test whether our results are consistent with a picture wherein massive z$\simeq$7 systems often reside in large ionized bubbles.
Given the rarity of such luminous high-redshift galaxies, we select these sources over very wide-area fields, specifically the COSMOS and XMM3 fields which span 1.5 and 1.8 deg$^2$, respectively.
We describe our z$\simeq$7 and z$\simeq$6 source selection criteria in turn below.

\subsubsection{Selection of z$\simeq$7 Sources} \label{sec:z7_selection}

To ensure we can reliably identify z$\simeq$7 galaxies and infer their \OIIIHb{} EWs, we photometrically select our targets over wide-area ($>$1 deg$^2$) fields with deep optical through mid-infrared imaging.
Because of the exceptionally deep imaging over COSMOS, the majority of our sources are selected within this field.
The Subaru/Hyper Surpime-Cam (HSC) Subaru Strategic Program (HSCSSP; \citealt{Aihara2018,Aihara2019}) provides optical (0.4--1$\mu$m) imaging across COSMOS in the \textit{g},\textit{r},\textit{i},\textit{z},\textit{y} broad-bands as well as the nb816 and nb921 narrow-bands.
The near-infrared data come from the UltraVISTA survey \citep{McCracken2012} which delivers imaging in VISTA/VIRCam \textit{Y}, \textit{J}, \textit{H}, and \textit{K}$_s$ broad-bands. 
We use PDR2 and DR4 data from HSCSSP and UltraVISTA, respectively, both of which are already astrometrically calibrated to the Gaia frame which we adopt throughout this work.

Our z$\simeq$7 source selection over COSMOS largely follows that described in \citetalias{Endsley2021} which utilizes, in part, the nb921 photometry to limit the selection to z$\gtrsim$6.6 with high confidence.
The specific adopted Lyman-break colour cuts are \textit{z}$-$\textit{y}$>$1.5, \textit{z}$-$\textit{Y}$>$1.5, nb921$-$\textit{Y}$>$1.0, and \textit{y}$-$\textit{Y}$<$0.4 where we set fluxes in the \textit{z} and nb921 dropout filters to their 1$\sigma$ values in cases of non-detections, consistent with previous literature \citep[e.g.][]{Bouwens2015_LF,Stefanon2019}.
These z$\simeq$7 selection criteria were designed by simulating the colours of z=6--8 galaxies with flat rest-UV slopes ($\beta$ = $-$2 where F$_{\nu} \propto \lambda^{\beta+2}$) using the \citet{Inoue2014} IGM transmission function.
As with any Lyman-break selection utilizing broad-band photometry, the exact redshift selection window depends slightly on the assumed \Lya{} EW.
Galaxies with weak \Lya{} emission ($\simeq$0 \AA{} EW) are only likely to satisfy the above criteria from z$\simeq$6.6--6.9 because at higher redshifts the \textit{y}$-$\textit{Y} colour becomes too red.
On the other hand, sources with strong \Lya{} satisfy our selection criteria to larger maximum redshifts -- up to z$\simeq$7.1 and z$\simeq$7.2 for \Lya{} EW = 25 \AA{} and 100 \AA{}, respectively. 
This is because strong \Lya{} will boost the \textit{y}-band flux resulting in bluer \textit{y}$-$\textit{Y} colours at a given redshift.
We refer the interested reader to Figure 2 of \citetalias{Endsley2021} for a visualization of the impact of \Lya{} on our redshift selection window.
Because only $\sim$10\% of bright z$\simeq$7 galaxies show strong \Lya{} emission (EW$>$25 \AA{}; \citealt{Ono2012,Schenker2014,Pentericci2018}), the large majority of our sample will likely lie at z$\simeq$6.6--6.9. 
None the less, we directly account for how our redshift selection window depends on \Lya{} strength when inferring the \Lya{} EW distribution as described in \S\ref{sec:completeness}.

Along with the above colour cuts, we ensure that each z$\simeq$7 source is real by  requiring a $>$3$\sigma$ detection in \textit{y}, \textit{Y}, and \textit{J} as well as a $>$5$\sigma$ detection in at least one of those three bands.
We also enforce non-detections ($<$2$\sigma$) in \textit{g} and \textit{r} since both these bands probe flux blueward of the Lyman-continuum limit at z$\geq$6.6. 
Finally, we clean our sample of T-type brown dwarfs (which can exhibit very strong \textit{z} and nb921-drops) by preserving only those with \textit{Y}$-$\textit{J}$<$0.45 or (\textit{J}$-$\textit{H}$>$0 and \textit{J}$-$\textit{K}$_s$~$>$0). 
These cuts are guided by brown dwarf spectra in the SPEX library \citep{Burgasser2014_SPEX} which demonstrate that T-type brown drafts have red \textit{Y}$-$\textit{J} colours but blue \textit{J}$-$\textit{H} and \textit{J}$-$\textit{K}$_s$ colours.

To build statistics on the brightest end, we also select galaxies over the XMM3 field. 
While the optical through mid-infrared coverage over XMM3 is significantly shallower relative to COSMOS, we are still able to identify the most luminous (\textit{J} $\lesssim$ 24.5) z$\simeq$7 galaxies within this field.
Near-infrared imaging over XMM3 comes from DR3 of the VIDEO survey\footnote{We astrometrically correct the public VIDEO mosaics to the Gaia frame to bring into agreement with the optical HSC imaging.} \citep{Jarvis2013} which provides data from the \textit{Z}, \textit{Y}, \textit{J}, \textit{H}, and \textit{K}$_s$ VISTA/VIRCam broad-bands.
For the optical imaging, we again use PDR2 of the HSCSSP which provides coverage in all the same bands as in COSMOS except for nb921. 
To compensate for the lack of nb921 imaging over XMM3, we enforce a much stronger dropout in the \textit{Z} band, specifically requiring \textit{Z}$-$\textit{y}$>$2.5 and \textit{Z}$-$\textit{Y}$>$2.5 while maintaining the \textit{y}$-$\textit{Y}$<$0.4 cut we applied in COSMOS.
We found that such cuts select z$\simeq$7 sources in a very similar redshift window as COSMOS after performing colour simulations of z=6--8 galaxies similar to those described in \citetalias{Endsley2021}.
Here, we use VIRCam \textit{Z} as the dropout band in XMM3 because the imaging in \textit{Z} is often slightly deeper relative to \textit{z} and both bands have very similar (normalized) transmission curves. 
All other selection criteria are equivalent to that over COSMOS.

To identify sources within each field, we run \textsc{SExtractor} \citep{Bertin1996} on a \textit{yYJHK$_s$} $\chi^2$ detection image \citep{Szalay1999}.
We then calculate the optical and near-infrared photometry of each source in 1.2\arcsec{} diameter apertures which is $\approx$1.5$\times$ the seeing in all bands.
Aperture corrections are calculated from the median curve-of-growth of nearby isolated, unsaturated stars and photometric errors are determined using the standard deviation of flux within apertures randomly placed in nearby empty locations.
As reported in \citetalias{Endsley2021}, the typical 5$\sigma$ depths in the COSMOS field are  m=27.7, 27.4, 27.3, 26.1, 26.9, 26.2, and 26.1 in \textit{g}, \textit{r}, \textit{i}, nb816, \textit{z}, nb921, and \textit{y}, respectively.
For the UltraVISTA data, the typical depths are m=24.7, 24.5, 24.2, and 24.5 in \textit{Y}, \textit{J}, \textit{H}, and \textit{K$_s$}, respectively, for the deep stripes and m=25.9, 25.8, 25.6, and 25.2, respectively for the ultra-deep stripes.
We are therefore able to identify z$\simeq$7 galaxies as faint as \textit{J}$\sim$26 in the ultra-deep stripes and \textit{J}$\sim$25 in the deep stripes.
In XMM3, the typical 5$\sigma$ depths are m=26.7, 26.3, 25.7, 25.3, 25.5, 24.3, 25.6, 25.0, 24.7, 24.3, and 23.9 in \textit{g}, \textit{r}, \textit{i}, nb816, \textit{z}, \textit{y}, \textit{Z}, \textit{Y}, \textit{J}, \textit{H}, and \textit{K}$_s$, respectively. 
We can therefore identify z$\simeq$7 galaxies as faint as \textit{J}$\sim$24.5 across XMM3.

\subsubsection{Selection of z$\simeq$6 Sources} \label{sec:z6_selection}

To quantify the evolution in the \Lya{} transmission for massive galaxies between z$\simeq$6--7, we also observed bright z$\simeq$6 sources that were photometrically selected as follows.
Similar to our z$\simeq$7 procedure, we design optimal z$\simeq$6 selection criteria by simulating the HSC and VIRCam colours of z=5--7 galaxies with rest-UV slope $\beta$ = $-$2 and adopting the IGM transmission function from \citet{Inoue2014}. 

Our resulting z$\simeq$6 selection criteria over COSMOS includes the following colour cuts:
\begin{enumerate}
    \item nb816$-$\textit{z}$>$1.5
    \item \textit{r}$-$\textit{z}$>$1.5
    \item \textit{z}$-$\textit{y}$<$0.5
    \item $-$0.5$<$nb921$-$\textit{y}$<$0.5
\end{enumerate}
Similar to our z$\simeq$7 selection, fluxes in nb816 and \textit{r} are set to their 1$\sigma$ value in cases of non-detections.
With these criteria, most selected bright z$\simeq$6 sources (i.e. those with weak \Lya{}; \citealt{Stark2011,deBarros2017}) will lie at z$\simeq$5.75--6.25 while strong \Lya{} emitters can be selected up to z$\simeq$6.5.
We have enforced the cut $-$0.5$<$nb921$-$\textit{y}$<$0.5 to minimize the likelihood that strong z$>$6.5 \Lya{} emitters are scattering into our z$\simeq$6 selection.
We additionally require a $>$5$\sigma$ detection in \textit{z} as well as a $>$3$\sigma$ detection in nb921 and \textit{y} to ensure each source is real.
Finally, we enforce non-detections ($<$2$\sigma$) in \textit{g} given that this band lies blueward of the Lyman-continuum break at z$>$5.

In XMM3, our z$\simeq$6 selection is largely similar. 
Specifically, we enforce nb816$-$\textit{Z}$>$1.5, \textit{r}$-$\textit{Z}$>$1.5, and \textit{Z}$-$\textit{Y}$<$0.5 (as detailed in \S\ref{sec:z7_selection}, the \textit{Z} and \textit{Y} imaging are generally deeper than \textit{z} and \textit{y}, respectively, across XMM3).
We also require a $>$5$\sigma$ detection in \textit{Z} as well as a $>$3$\sigma$ detection in \textit{z} and \textit{Y} to ensure each source is real, in addition to non-detections ($<$2$\sigma$) in \textit{g}.
As in COSMOS, these cuts result in a redshift selection window of z$\simeq$5.75--6.25 for sources with weak \Lya{} emission (EW$\simeq$0 \AA{}).
Due to the lack of nb921 imaging across this field, sources with very strong \Lya{} emission (EW$\sim$100 \AA{}) can be selected up to z$\simeq$6.6.
However, our spectroscopic data confirm that all three of our z$\simeq$6 XMM3 targets lie at z$<$6.5 (\S\ref{sec:binospec}) so this is not a concern for this work.

These z$\simeq$6 sources are identified after running \textsc{SExtractor} on a \textit{z}nb921\textit{yYJHK}$_s$ and \textit{zZyYJHK}$_s$ $\chi^2$ detection image over COSMOS and XMM3, respectively. 
Optical and near-infrared photometry are calculated in the same way as for the z$\simeq$7 sources (1.2\arcsec{} diameter apertures). 
Given the depths in each field, we are able to identify z$\simeq$6 sources down to \textit{y}$\sim$26.5 in COSMOS\footnote{Because our z$\simeq$6 selection criteria in COSMOS does not utilize VIRCam photometry, the selection is equivalent for the deep and ultra-deep UltraVISTA stripes.} and \textit{Y}$\sim$25.5 in XMM3. 
Note that throughout this work, for the z$\simeq$6 galaxies, we quote \textit{y}-band magnitudes for those in COSMOS and \textit{Y}-band magnitudes for those in XMM3 given the depth differences noted above.

\begin{table*}
\centering
\caption{Summary of our MMT/Binospec observations.}
\begin{tabular}{P{1.5cm}P{1.8cm}P{1.8cm}P{1.2cm}P{1.2cm}P{1.8cm}P{1.8cm}} 
\hline
Mask Name & RA & Dec & PA & $\lambda_{\mathrm{cen}}$ & Exposure Time & Average Seeing \Tstrut \\
 & &  & [deg] & [\AA{}] & [s] & [arcsec] \Bstrut \\ 
\hline
COSa & 10:02:29.53 & $+$02:17:24.58 & $+$45.0 & 8500 & 7200 & 1.0 \Tstrut{} \\
COSb & 10:00:28.37 & $+$01:53:50.84 & $-$98.5 & 8700 & 32400 & 1.1 \\
COSc & 09:59:09.41 & $+$02:21:25.80 & $-$101.0 & 8700 & 7200 & 1.1 \\
COSd & 09:59:10.01 & $+$02:21:07.83 & $-$101.0 & 8700 & 18900 & 1.0 \\
COSe & 10:00:39.92 & $+$02:35:16.12 & $-$20.0 & 8720 & 17100 & 1.2 \\
XMM3a & 02:26:38.04 & $-$05:03:11.15 & $-$12.0 & 8700 & 10200 & 0.9 \\
XMM3b & 02:27:22.70 & $-$04:19:35.44 & $-$116.2 & 8700 & 18900 & 1.0 \\
\hline
\end{tabular}
\label{tab:binomask_info}
\end{table*}

\begin{table*}
\centering
\caption{z$\simeq$7 galaxies targeted with MMT/Binospec across the $\approx$1.5 deg$^2$ and 1.8 deg$^2$ COSMOS and XMM3 fields, respectively. For sources with a non-detection (S/N$<$1) in one of the IRAC bands, we report the 2$\sigma$ limiting magnitude and colour.} \label{tab:z7_phot_properties}
\begin{tabular}{P{2.0cm}P{1.2cm}P{1.45cm}P{1.3cm}P{1.3cm}P{1.3cm}P{1.3cm}P{2.3cm}}
\hline
Source ID & RA & Dec & \textit{J} & 3.6 $\mu$m & 4.5 $\mu$m & [3.6]$-$[4.5] & Masks \Tstrut\Bstrut \\[2pt]
\hline
COS-221419 & 10:00:26.28 & +01:46:03.22 & 26.07$^{+0.29}_{-0.23}$ & 25.22$^{+0.25}_{-0.21}$ & $>$26.30 & $<$-1.08 & COSb \Tstrut{} \\[4pt]
COS-235129 & 10:00:39.21 & +01:46:43.68 & 25.75$^{+0.26}_{-0.20}$ & 24.94$^{+0.21}_{-0.18}$ & 25.03$^{+0.16}_{-0.14}$ & -0.09$^{+0.26}_{-0.24}$ & COSb \\[4pt] 
COS-237729 & 10:00:31.42 & +01:46:51.01 & 25.68$^{+0.18}_{-0.15}$ & 24.84$^{+0.20}_{-0.17}$ & 25.33$^{+0.23}_{-0.19}$ & -0.49$^{+0.28}_{-0.28}$ & COSb \\[4pt] 
COS-301652 & 10:00:54.82 & +01:50:05.18 & 25.65$^{+0.22}_{-0.18}$ & 24.38$^{+0.11}_{-0.10}$ & 24.67$^{+0.13}_{-0.11}$ & -0.28$^{+0.16}_{-0.16}$ & COSb \\[4pt] 
COS-469110 & 10:00:04.36 & +01:58:35.53 & 24.97$^{+0.30}_{-0.23}$ & 24.28$^{+0.10}_{-0.09}$ & 24.69$^{+0.17}_{-0.14}$ & -0.40$^{+0.18}_{-0.19}$ & COSb \\[4pt] 
COS-505871 & 10:00:21.35 & +02:00:30.93 & 25.51$^{+0.16}_{-0.14}$ & 24.39$^{+0.09}_{-0.09}$ & 24.54$^{+0.13}_{-0.12}$ & -0.16$^{+0.15}_{-0.16}$ & COSb \\[4pt] 
COS-534584 & 10:00:42.13 & +02:01:56.87 & 24.99$^{+0.12}_{-0.11}$ & 24.02$^{+0.10}_{-0.09}$ & 24.44$^{+0.14}_{-0.13}$ & -0.42$^{+0.16}_{-0.17}$ & COSb \\[4pt] 
COS-788571 & 09:59:21.68 & +02:14:53.02 & 25.27$^{+0.11}_{-0.10}$ & 24.40$^{+0.09}_{-0.08}$ & 25.32$^{+0.23}_{-0.19}$ & -0.92$^{+0.21}_{-0.24}$ & COSd \\[4pt] 
COS-851423 & 09:59:11.46 & +02:18:10.42 & 25.91$^{+0.22}_{-0.19}$ & 24.82$^{+0.15}_{-0.13}$ & 25.54$^{+0.47}_{-0.32}$ & -0.72$^{+0.36}_{-0.48}$ & COSc \& COSd \\[4pt] 
COS-854905 & 09:59:09.13 & +02:18:22.38 & 25.75$^{+0.28}_{-0.22}$ & 24.46$^{+0.19}_{-0.16}$ & 24.90$^{+0.32}_{-0.25}$ & -0.44$^{+0.31}_{-0.35}$ & COSc \& COSd \\[4pt] 
COS-856875 & 09:58:45.34 & +02:18:28.87 & 25.64$^{+0.30}_{-0.24}$ & 25.09$^{+0.25}_{-0.20}$ & 25.66$^{+0.39}_{-0.28}$ & -0.57$^{+0.38}_{-0.44}$ & COSc \\[4pt] 
COS-862541 & 10:03:05.25 & +02:18:42.75 & 24.49$^{+0.26}_{-0.21}$ & 23.33$^{+0.09}_{-0.08}$ & 24.65$^{+0.30}_{-0.24}$ & -1.33$^{+0.26}_{-0.32}$ & COSa \\[4pt] 
COS-940214 & 09:59:06.73 & +02:22:45.93 & 26.27$^{+0.43}_{-0.31}$ & 25.06$^{+0.29}_{-0.23}$ & $>$26.32 & <-1.26 & COSd \\[4pt] 
COS-955126 & 09:59:23.62 & +02:23:32.73 & 25.38$^{+0.24}_{-0.20}$ & 24.20$^{+0.14}_{-0.13}$ & 25.14$^{+0.43}_{-0.30}$ & -0.94$^{+0.33}_{-0.44}$ & COSd \\[4pt] 
COS-1009842 & 09:59:06.33 & +02:26:30.48 & 26.22$^{+0.25}_{-0.20}$ & 25.16$^{+0.24}_{-0.20}$ & 25.78$^{+0.49}_{-0.34}$ & -0.61$^{+0.42}_{-0.52}$ & COSc \& COSd \\[4pt] 
COS-1048848 & 09:59:09.76 & +02:28:32.95 & 26.09$^{+0.27}_{-0.22}$ & 26.11$^{+0.65}_{-0.40}$ & $>$26.24 & <-0.13 & COSc \& COSd \\[4pt] 
COS-1053257 & 09:58:46.20 & +02:28:45.76 & 24.79$^{+0.08}_{-0.07}$ & 23.81$^{+0.23}_{-0.19}$ & 24.13$^{+0.30}_{-0.24}$ & -0.33$^{+0.33}_{-0.36}$ & COSc \& COSd \\[4pt] 
COS-1099982 & 10:00:23.37 & +02:31:14.80 & 25.45$^{+0.14}_{-0.13}$ & 24.11$^{+0.09}_{-0.09}$ & 25.43$^{+0.26}_{-0.21}$ & -1.32$^{+0.23}_{-0.28}$ & COSe \\[4pt] 
COS-1205190 & 10:00:45.44 & +02:36:48.81 & 25.82$^{+0.20}_{-0.17}$ & 25.81$^{+0.89}_{-0.48}$ & $>$25.82 & $<$-0.01 & COSe \\[4pt] 
COS-1235751 & 10:00:11.57 & +02:38:29.81 & 25.62$^{+0.22}_{-0.18}$ & 24.27$^{+0.14}_{-0.12}$ & 24.45$^{+0.16}_{-0.14}$ & -0.18$^{+0.19}_{-0.20}$ & COSe\\[4pt] 
XMM3-227436 & 02:26:46.19 & -04:59:53.57 & 24.67$^{+0.21}_{-0.18}$ & 24.37$^{+0.30}_{-0.23}$ & 23.78$^{+0.18}_{-0.16}$ & 0.58$^{+0.33}_{-0.30}$ & XMM3a \\[4pt] 
XMM3-504799 & 02:27:13.12 & -04:17:59.25 & 24.33$^{+0.16}_{-0.14}$ & 23.37$^{+0.11}_{-0.10}$ & 24.32$^{+0.41}_{-0.30}$ & -0.95$^{+0.32}_{-0.43}$ & XMM3b \\[4pt] 
\hline
\end{tabular}
\end{table*}

\subsubsection{IRAC Photometry} \label{sec:IRAC}

To infer the [OIII]$+$H$\beta$ EWs of our z$\simeq$7 sample, we take advantage of the \Spitzer{}/IRAC imaging over both COSMOS and XMM3.
A full description of our procedure for generating IRAC mosaics in both the 3.6$\mu$m and 4.5$\mu$m filters is provided in \citetalias{Endsley2021}.
Briefly, we use the \textsc{mopex} software \citep{Makovoz2005_MOPEX} to coadd background-subtracted images (using \textsc{SExtractor}) and astrometrically match the output mosaics to the Gaia reference frame using the IRAF package \textsc{ccmap}. 
In COSMOS, the IRAC data comes from a multitude of surveys: the \Spitzer{} Extended Deep Survey (SEDS; \citealt{Ashby2013}), S-CANDELS \citep{Ashby2015}, Star Formation at 4$<$z$<$6 from the \Spitzer{} Large Area Survey with Hyper Suprime-Cam (SPLASH; \citealt{Steinhardt2014}), \Spitzer{} Matching survey of the UltraVISTA ultra-deep Stripes (SMUVS; \citealt{Ashby2018}), and Completing the Legacy of \Spitzer{}/IRAC over COSMOS (P.I. I. Labb\'e).
The IRAC data over XMM3 largely comes from the \Spitzer{} Extragalactic Representative Volume Survey (SERVS; \citealt{Mauduit2012}) with deeper imaging on our z$\simeq$7 targets from late-2019 observations led by P.I. M. Stefanon. 

To compensate for the considerably broader IRAC PSF relative to the optical/near-infrared seeing, we measure IRAC photometry in 2.8\arcsec{} diameter apertures and utilize a deconfusion algorithm to remove contaminating flux from neighboring sources.
In XMM3, our deconfusion approach is equivalent to that detailed in \citetalias{Endsley2021}.
To summarize, we convolve the flux profile of every nearby source detected in the \textit{yYJHK}$_s$ $\chi^2$ image with a 2D Gaussian having FWHM equal to the quadrature difference of the IRAC FWHM and the median seeing from each band in our $\chi^2$ detection images.
Flux profiles are calculated as the square root of the $\chi^2$ image using the \textsc{SExtractor} segmentation map to determine source footprints.
The convolved flux profiles are fit to the IRAC image with total fluxes of each source as free parameters.
Once this is done, the best-fitting flux profile of each neighboring source is subtracted before measuring the IRAC photometry.
We note that our XMM3 targets are not strongly confused and that residuals from the deconfusion algorithm are acceptably smooth.
That is the S/N of the pixels inside the aperture do not appear to be systematically offset due to poor flux-profile fitting of the neighboring sources upon visual inspection.
We show the deconfused IRAC postage stamps of both sources in Fig. \ref{fig:IRAC} (see Appendix) to illustrate this point.

In COSMOS, we take advantage of the very high-resolution \HST{} F814W imaging across this field \citep{Scoville2007_HST} to calculate the flux profile of each neighboring source.
By convolving these flux profiles with IRAC PSFs calculated using unsaturated stars near each source ($<$3\arcmin{} separation), we are able to obtain much smoother residual images from the deconfusion algorithm for sources lying in crowded regions.
For the purposes of this work, we remove sources from our sample that have poor IRAC residuals after employing our deconfusion algorithm, as is common in the literature \citep[e.g.][]{Labbe2013,Smit2015,Bouwens2015_LF,deBarros2019}.
One of our COSMOS targets (COS-862541) lies outside the F814W imaging and we therefore employ the same deconfusion approach as for XMM3, noting that this source is also not strongly confused.
The deconfused IRAC postage stamp images of all our targets are shown in Fig. \ref{fig:IRAC} (see Appendix).

\subsection{MMT/Binospec Spectroscopy} \label{sec:binospec}

We have followed up a subset of our z$\simeq$6 and z$\simeq$7 galaxy samples described in \S\ref{sec:sample_selection} using the Binospec spectrograph \citep{Fabricant2019} installed at the MMT. 
Binospec is a wide-field (240 arcmin$^2$) and multi-object (up to $\sim$150 sources) spectrograph enabling wavelength coverage up to $\approx$1$\mu$m with moderately high resolution ($R \approx 4400$).
As such, Binospec is an ideal instrument to target \Lya{} in large numbers of z$\simeq$6--7 galaxies selected over wide-area fields. 
So far, we have observed seven masks with five in COSMOS and two in XMM3, totaling just over 31 hours of integration time across all masks. 
In Table \ref{tab:binomask_info}, we report the central coordinate, position angle, central wavelength, total exposure time, and average seeing for each mask.
When choosing which sources would be assigned slits for each mask, we assigned higher priority to sources with brighter UV magnitudes to obtain a more homogeneous sample across luminosity.
Because galaxy rest-UV slopes typically become bluer at higher redshifts \citep[e.g.][]{Bouwens2014_beta}, we also gave higher priority to z$\simeq$6 sources with bluer rest-UV slopes to obtain z$\simeq$6 and z$\simeq$7 samples with more similar properties.

\begin{figure}
\includegraphics{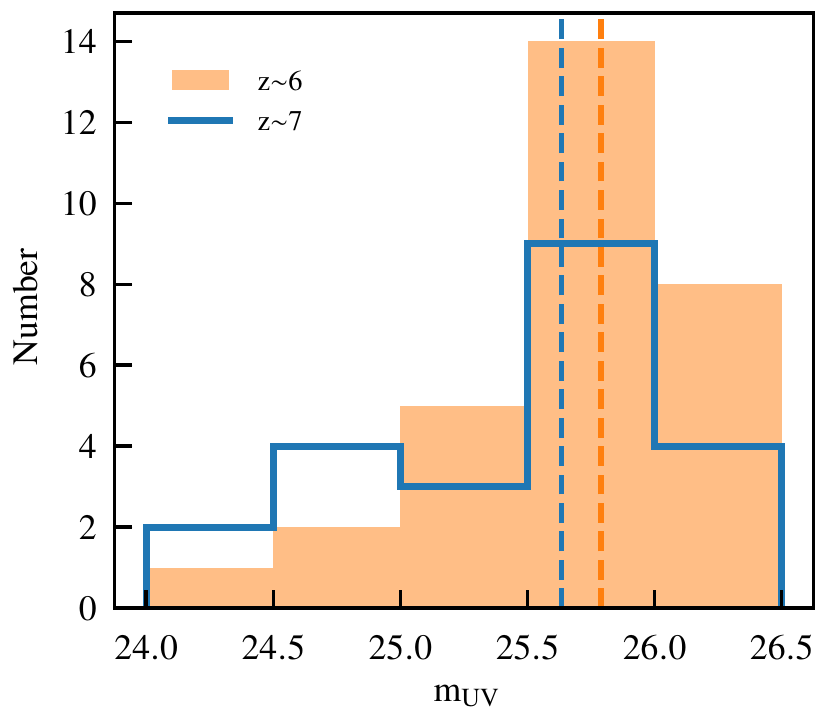}
\includegraphics{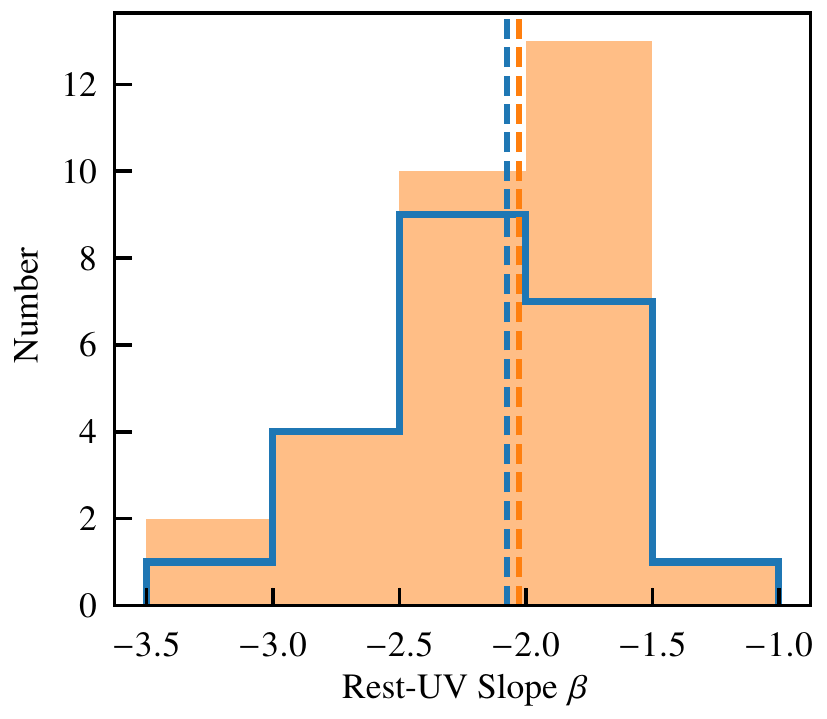}
\caption{Figures comparing the distribution of apparent UV magnitudes (top) and rest-UV slopes (bottom) for our spectroscopic z$\simeq$6 and z$\simeq$7 samples. We show the z$\simeq$6 distributions with a shaded orange histogram while the z$\simeq$7 distributions are shown with a thick blue line. We show median values for each sample with vertical dashed lines. The range of UV magnitudes and rest-UV slopes, as well as their typical values, are very similar for these two spectroscopic samples.}
\label{fig:sampleComparison}
\end{figure}

\subsubsection{Sample Description} \label{sec:observed_samples}

Across these seven masks, we have targeted 22 z$\simeq$7 sources satisfying the selection criteria described in \S\ref{sec:sample_selection}.
In Table \ref{tab:z7_phot_properties}, we report the coordinates, \textit{J}-band magnitudes, 3.6 and 4.5$\mu$m photometry, as well as the [3.6]$-$[4.5] colour for each targeted z$\simeq$7 galaxy.
We also note in Table \ref{tab:z7_phot_properties} the mask containing each source, where four sources were placed on both the COSc and COSd masks for a total integration time of 26100 seconds.
The \textit{J}-band magnitudes of our z$\simeq$7 targets range from \textit{J} = 24.3--26.3.
The vast majority of these sources show blue [3.6]$-$[4.5] colours ($<$0), as expected given that z$\gtrsim$7 galaxies typically possess strong \OIIIHb{} emission (\citealt{Labbe2013}; \citealt{Smit2014}; \citealt{deBarros2019}; \citetalias{Endsley2021}).
Furthermore, 8 of our 22 z$\simeq$7 targets have very blue [3.6]$-$[4.5] colours ($<$ $-$0.8) which translates to strong \OIIIHb{} emission (EW$\gtrsim$800 \AA{}) assuming a flat rest-optical continuum slope and z=6.75. 
This includes two sources (COS-221419 and COS-940214) that have 3.6$\mu$m fluxes substantially above the 4.5$\mu$m 2$\sigma$ upper limit.
Thus, even though these sources are not detected in 4.5$\mu$m, the data still suggest that they likely possess strong \OIIIHb{} emission.

Within these same seven masks, we have also targeted 30 z$\simeq$6 galaxies satisfying the selection criteria from \S\ref{sec:sample_selection}.
The rest-UV apparent magnitudes\footnote{As discussed in \S\ref{sec:z6_selection}, we use \textit{y} and \textit{Y}-band magnitudes to quote \muv{} for z$\simeq$6 sources in COSMOS and XMM3, respectively.} of these z$\simeq$6 targets range from \muv{} = 24.3--26.4, very similar to that of the z$\simeq$7 targets (Fig. \ref{fig:sampleComparison} top).
The range of rest-UV slopes spanned by our z$\simeq$6 spectroscopic targets ($-1.2 \leq \beta \leq -3.5$) is also very similar to that of the z$\simeq$7 targets ($-1.1 \leq \beta \leq -3.4$).
Furthermore, the typical rest-UV slopes of each sample are almost exactly equal -- the median $\beta$ of the z$\simeq$7 sample is $-$2.07 and that of the z$\simeq$6 sample is $-$2.09 (Fig. \ref{fig:sampleComparison} bottom).
For the z$\simeq$7 sources, we calculate rest-UV slopes using \textit{YJHK}$_s$, while for the z$\simeq$6 sources we use nb921\textit{yYJHK}$_s$ and \textit{yYJHK}$_s$ in COSMOS and XMM3, respectively.

\subsubsection{Data Reduction} \label{sec:reduction}

To design our Binospec multi-object slit masks, we used the \textsc{BinoMask} software.
We adopted a slit width of 1.0\arcsec{} and the 600 l/mm grating, yielding a resolving power of $R \approx 4360$.
Central wavelengths for each mask range from 8500--8720 \AA{} (see Table \ref{tab:binomask_info}) with the exact value chosen to optimize the red-end wavelength coverage of our z$\simeq$7 targets.
For all z$\simeq$7 targets, the maximum wavelength coverage was $\geq$9635 \AA{} meaning that our observations covered \Lya{} up to at least \zLya{} = 6.93.
Furthermore, the typical wavelength coverage of our z$\simeq$7 targets was 0.75--1.00$\mu$m (\zLya{} = 5.17--7.23).
For the z$\simeq$6 targets, our observations covered at least 7890--9510 \AA{} and therefore (assuming they all lie at z$>$5) fully encompassed the \Lya{} redshift range \zLya{} $\simeq$ 5.75--6.50 set by our selection criteria (\S\ref{sec:z6_selection}).

We adopted a slit length of at least 20\arcsec{} (7\arcsec{}) for the z$\simeq$7 (z$\simeq$6) targets which we found leads to sufficient modeling of the sky background from $\approx$0.9--1$\mu$m ($\approx$0.8--0.9$\mu$m).
Each mask also included at least five stars off the primary field of views\footnote{Binospec's field of view is composed of two separate 8$\times$15 arcmin$^2$ FoVs separated by a 3.2 arcminute gap.} (FoVs) for guiding and monitoring seeing throughout the observations.
In addition, we placed three stars within the primary FoVs to determine exposure weighting (see below) and absolute flux calibration.

Each individual exposure was reduced separately using the publicly available Binospec data reduction pipeline \citep{Kansky2019}. 
Exposures for each mask were then co-added using the weighting scheme from \citet{Kriek2015} which takes the height of the telluric-corrected 1D flux profile of bright stars on the mask as the relative weight.
This accounts for both the average sky transmission and relative seeing in each exposure.
The average seeing of each mask (calculated from the co-added spectra) is listed in Table \ref{tab:binomask_info} and ranges from 0.9--1.2\arcsec{}.

We extract 1D spectra using optimal extraction \citep{Horne1986} after fitting a Gaussian\footnote{The FWHM of this Gaussian is not allowed to be less than the seeing.} to the observed emission line profile along the spatial axis. 
As in \citetalias{Endsley2021}, absolute flux calibration is determined by calculating the average scaling factor that matches the 1D spectra of the three stars placed on the mask to their mean PSF $z$-band magnitudes from the Pan-STARRS survey \citep{Chambers2016_PanSTARRS}. 
Given the relatively narrow wavelength range covered by these observations ($\approx$0.75-1$\mu$m), we assume that this factor does not evolve with wavelength.

To estimate slit loss factors, we adopt the size-luminosity relation of bright z=6--7 galaxies found by \citet{CurtisLake2016} which assumes a S\'ersic profile with $n = 1.0$ (consistent with the approach of \citealt{Ono2013}). 
The modeled S\'ersic profile of each source is convovled with a 2D Gaussian with FWHM equal to the seeing of the respective mask, and the fraction of flux within the 1\arcsec{} Binospec slit is compared to that for a point source. 
The resulting relative slit loss correction factors range from 4--10\% for all z$\simeq$6 and z$\simeq$7 targets, with brighter sources having larger corrections due to their larger assumed half-light radius.
We note that our results are not significantly altered if we instead calculate slit loss correction factors assuming the size-luminosity relation from \citet{Bowler2017} which mainly yields larger sizes (and hence larger correction factors) for the brightest (\muv{} $<$ 25) sources.

We identify emission lines by first visually inspecting the 2D spectra of all sources.
For emission features relatively clear of strong skylines, we calculate the flux in an aperture where the width along the wavelength axis is set by visual inspection of the 2D spectra with flux errors computed as described in \citetalias{Endsley2021}.
For these relatively clean emission features, we estimate that the fraction\footnote{This fraction is calculated by assuming a fixed \Lya{} line profile shape equivalent to that used in our completeness simulations (Gaussian with FWHM = 220 km s$^{-1}$; \S\ref{sec:completeness}) and fitting this profile to the 1D spectra of each detected source.} of flux potentially obscured by strong skylines is small (15--20\%) on average in both the z$\simeq$6 and z$\simeq$7 galaxies. 
We therefore conclude that our results on the evolution of the \Lya{} EW distribution between these two redshifts (\S\ref{sec:evolution}) are not significantly impacted by this slight obscuration. 
Such small obscuration would furthermore only serve to slightly strengthen our conclusions\footnote{This is because the \Lya{} EW for detected sources (which almost always lie in the strong \OIIIHb{} emitter sub-sample; \S\ref{sec:z7_LyA}) would be pushed to higher values.} on differences in the \Lya{} EW distribution for moderate versus strong \OIIIHb{} emitters at z$\simeq$7 (\S\ref{sec:z7_LyA}). 
Because the exact skyline obscuration fraction for a given source depends on the assumed line profile and accounting for it does not significantly alter our conclusions, we do not fold an obscuration correction for these relatively clean sources into our analysis.

For the few features that overlap more significantly with moderate-strength skylines, we adopt a slightly different approach to minimize the impact of skylines on the recovered flux (these sources are marked with asterisks in Tables \ref{tab:LyA_detection_info_z7} and \ref{tab:LyA_detection_info_z6}).
We fit the 1D spectra with a half-Gaussian (red-side only) convolved with the spectral resolution of our instrument \citep[e.g.][]{Hu2010} after masking portions of the spectrum contaminated by skylines.
In this fitting procedure, we adopt a grid of three parameters describing the half-gaussian (amplitude, standard deviation, and central wavelength) and compute the $\chi^2$ value for each point in the grid.
The likelihood for a given set of parameters is then calculated as $P(A,\sigma,\lambda_0) \propto e^{-\chi^2/2}$ which we convert to a probability distribution on the flux of the emission line. 
The flux and its uncertainty is then computed as the median value and standard deviation from this probability distribution.
For all emission features, we subtract the continuum flux estimated from the photometry\footnote{For z$\simeq$7 sources with \zLya{}$<$6.9, we use the \textit{Y}-band photometry to estimate the continuum. For those with \zLya{}$>$6.9, we use the \textit{J} band because \textit{Y} is partially contaminated by \Lya{} at these redshifts. For the z$\simeq$6 sources, we adopt the \textit{y} and \textit{Y}-band photometry for the continuum in COSMOS and XMM3, respectively.} but compute the significance of the feature prior to this subtraction.

\begin{figure}
\includegraphics{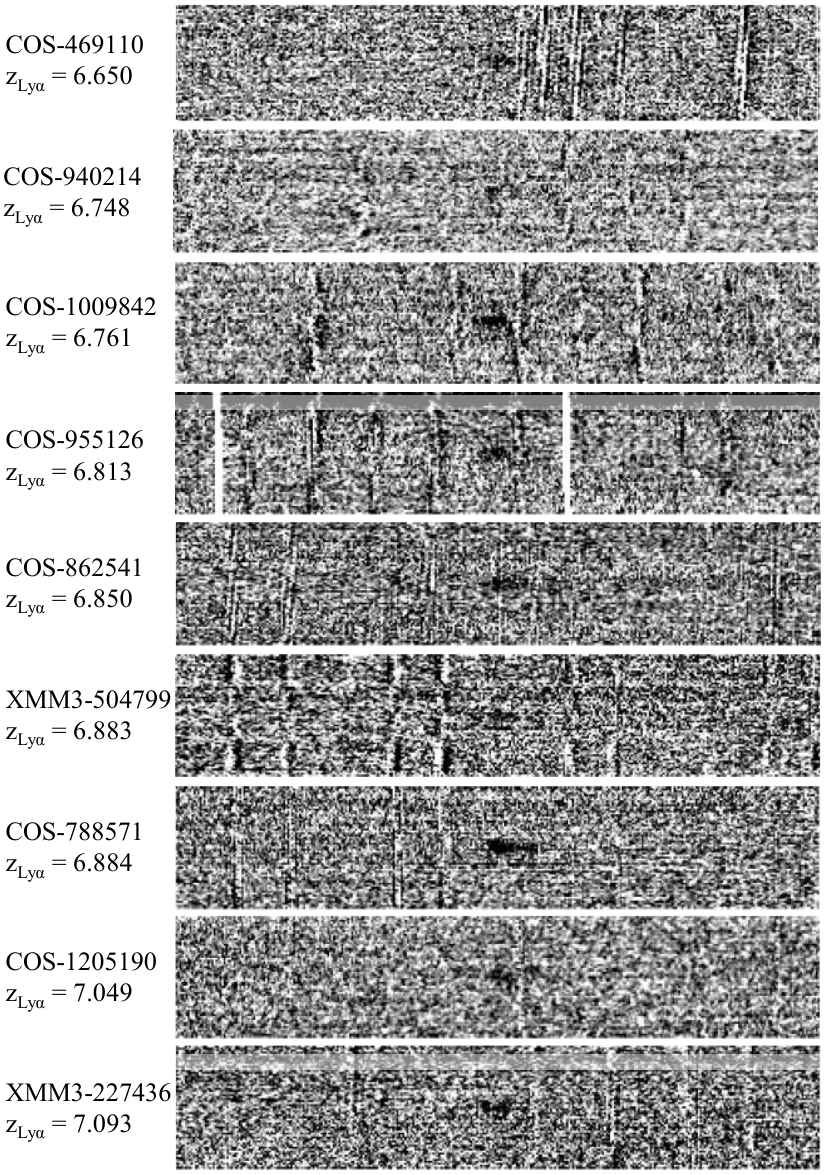}
\caption{2D signal-to-noise ratio maps (black is positive) of our nine confident ($>$7$\sigma$) \Lya{} detections at z$\simeq$7. These detections span a redshift range of \zLya{}=6.650--7.093. Each sub-figure spans $\pm$100 \AA{} along the x-axis and $\pm$7.3\arcsec{} along the y-axis. We mask the bright continuum from nearby sources in the spectra of COS-955126 and XMM3-227436 to improve clarity.}
\label{fig:LyADetectionMosaic_z7}
\end{figure}

\begin{table}
\centering
\caption{Information on confident ($>$7$\sigma$) \Lya{} detections in our z$\simeq$7 sample. We quote the significance of each detection within parenthesis in the flux column where the fluxes have the continuum subtracted. Sources marked with asterisks are those with emission features close to moderate-strength skylines where we used a line-profile fitting technique to derive their total fluxes.}
\begin{tabular}{P{2.0cm}P{0.7cm}P{2.3cm}P{1.2cm}} 
\hline
Source ID & \zLya{} & Flux & EW \Tstrut{} \\
 & & [10$^{-18}$ erg/s/cm$^2$] & [\AA{}] \Bstrut{}\\
\hline 
COS-469110 & 6.650 & 9.7$\pm$1.3 (7.8$\sigma$) & 14.5$\pm$5.0 \Tstrut{}\\[2pt]
COS-940214$^{\ast}$ & 6.748 & 11.8$\pm$1.8 (7.1$\sigma$) & 43.1$\pm$14.7 \\[2pt]
COS-1009842 & 6.761 & 12.3$\pm$0.8 (15.4$\sigma$) & 41.6$\pm$9.5 \\[2pt]
COS-955126 & 6.813 & 8.2$\pm$0.9 (10.4$\sigma$) & 12.3$\pm$2.5 \\[2pt]
COS-862541 & 6.850 & 15.3$\pm$1.9 (9.0$\sigma$) & 11.8$\pm$2.7 \\[2pt]
XMM3-504799 & 6.883 & 5.1$\pm$0.9 (7.1$\sigma$) & 3.7$\pm$0.8 \\[2pt]
COS-788571 & 6.884 & 16.3$\pm$1.1 (15.9$\sigma$) & 30.6$\pm$3.9 \\[2pt]
COS-1205190 & 7.049 & 12.4$\pm$1.6 (8.1$\sigma$) & 28.8$\pm$6.0 \\[2pt]
XMM3-227436 & 7.093 & 18.6$\pm$2.3 (8.9$\sigma$) & 15.0$\pm$3.2 \\[2pt]
\hline
\end{tabular}
\label{tab:LyA_detection_info_z7}
\end{table}

\subsubsection{Detected Emission Lines}

We detect confident ($>$7$\sigma$) emission features in 9 of the 22 z$\simeq$7 galaxies in our sample (see Fig. \ref{fig:LyADetectionMosaic_z7}).
In all cases, we interpret these features as \Lya{} because all are fully consistent with such a solution given the expected redshift range of our sample and none are consistent with an [OII]$\lambda$3727,3729 solution (two narrow peaks of roughly similar strength separated by $\approx$6.8 \AA{} in the observed frame).
We consider [OII]$\lambda$3727,3729 as the primary alternative solution because it would be very difficult for anything but a Balmer break to mimic the strong nb921 and \textit{Z} drops required by our selection criteria.
We also find no other convincing features in the spectra of these sources (aside from the tentative features in COS-469110 consistent with NV emission; see below).
The measured line fluxes and rest-frame EWs for sources with a \Lya{} detection are reported in Table \ref{tab:LyA_detection_info_z7}.
We also calculate the corresponding \Lya{} redshifts, \zLya{}, using the observed wavelength of peak flux in the 1D extraction and a rest-frame \Lya{} wavelength of 1215.67 \AA{}.
The detection for COS-862541 was previously reported in \citetalias{Endsley2021}.

Our z$\simeq$7 \Lya{} detections span redshifts of \zLya{} = 6.650--7.093. Measured fluxes and EWs range from (5.1--18.6)$\times$10$^{-18}$ erg/s/cm$^2$ and 3.7--43.1 \AA{}, respectively.
None of the detections lie at \zLya{}$<$6.6, consistent with the strong nb921 and \textit{Z} drops used in our selection.
Furthermore, the large majority of detected targets lie at \zLya{}$\simeq$6.6--6.9.
While two of our sources (COS-1205190 and XMM3-227436) lie at z=7.05--7.09, these redshifts are still consistent with our selection given the moderate-strength \Lya{} emission for these sources (EW = 15--29 \AA{}; see \S\ref{sec:z7_selection}). 

\begin{figure}
\includegraphics{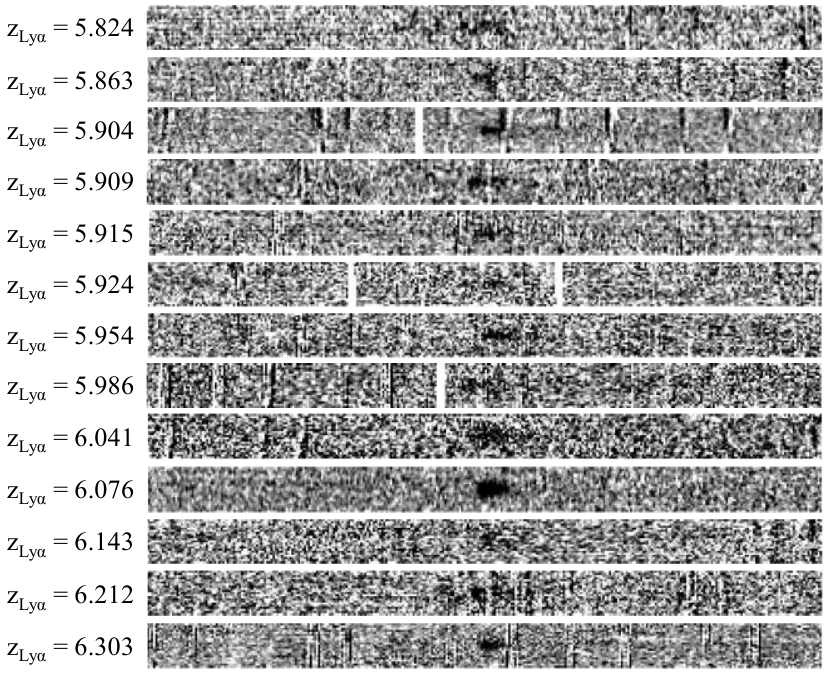}
\caption{2D signal-to-noise ratio maps of our 13 \Lya{} detections at z$\simeq$6 where black is positive. These detections span a redshift range of \zLya{}=5.824--6.303. Each sub-figure spans $\pm$100 \AA{} along the x-axis and $\pm$2.5\arcsec{} along the y-axis.}
\label{fig:LyADetectionMosaic_z6}
\end{figure}

\begin{table*}
\centering
\caption{Information on confident ($>$7$\sigma$) \Lya{} detections in our z$\simeq$6 sample. We report HSC \textit{y}-band magnitudes for sources in COSMOS and VIRCam \textit{Y}-band magnitudes in XMM3. Rest-UV slopes, $\beta$, are measured by fitting the nb921\textit{yYJHK}$_s$ in COSMOS and \textit{yYJHK}$_s$ in XMM3. We quote the significance of each detection within parenthesis in the flux column where the fluxes have the continuum subtracted. Sources marked with asterisks are those with emission features close to moderate-strength skylines where we used a line-profile fitting technique to derive their total fluxes.}
\begin{tabular}{P{2.0cm}P{1.4cm}P{1.65cm}P{1.5cm}P{1.5cm}P{0.7cm}P{2.3cm}P{1.2cm}} 
\hline
Source ID & RA & Dec & \textit{y}/\textit{Y} & $\beta$ & \zLya{} & Flux & EW \Tstrut{} \\
 & & & & & & [10$^{-18}$ erg/s/cm$^2$] & [\AA{}] \Bstrut{}\\
\hline 
XMM3-229059$^{\ast}$ & 02:26:22.67 & -05:05:31.13 & 24.31$^{+0.13}_{-0.11}$ & -2.29$\pm$0.19 & 5.824 & 27.6$\pm$4.8 (7.6$\sigma$) & 13.6$\pm$2.3\Tstrut{} \\[2pt]
COS-291078 & 10:00:41.08 & +01:47:18.54 & 25.73$^{+0.23}_{-0.19}$ & -1.95$\pm$0.39 & 5.863 & 10.9$\pm$1.3 (8.5$\sigma$) & 16.0$\pm$2.6 \\[2pt]
COS-1131140 & 09:59:20.27 & +02:23:22.22 & 25.63$^{+0.15}_{-0.13}$ & -1.66$\pm$0.12 & 5.904 & 6.8$\pm$0.7 (11.1$\sigma$) & 11.4$\pm$1.8 \\[2pt]
COS-1163498$^{\ast}$ & 09:59:18.45 & +02:24:53.93 & 26.21$^{+0.27}_{-0.22}$ & -1.81$\pm$0.30 & 5.909 & 10.2$\pm$1.3 (8.9$\sigma$) & 29.2$\pm$7.4 \\[2pt]
COS-905289 & 09:59:44.62 & +02:13:29.21 & 25.79$^{+0.20}_{-0.17}$ & -1.94$\pm$0.80 & 5.915 & 6.9$\pm$0.8 (9.0$\sigma$) & 13.4$\pm$2.7 \\[2pt]
COS-1181452 & 10:00:19.93 & +02:25:36.81 & 24.53$^{+0.08}_{-0.08}$ & -1.91$\pm$0.12 & 5.924 & 10.5$\pm$1.4 (8.9$\sigma$) & 6.4$\pm$0.9 \\[2pt]
COS-881759 & 09:58:56.89 & +02:12:29.64 & 26.20$^{+0.26}_{-0.20}$ & -2.03$\pm$0.37 & 5.954 & 9.1$\pm$0.8 (11.3$\sigma$) & 26.1$\pm$6.0 \\[2pt]
XMM3-569712$^{\ast}$ & 02:27:23.06 & -04:25:53.27 & 24.68$^{+0.15}_{-0.13}$ & -2.06$\pm$0.24 & 5.986 & 7.9$\pm$2.0 (7.1$\sigma$) & 5.5$\pm$1.0 \\[2pt]
COS-282685 & 10:00:55.03 & +01:46:56.00 & 25.49$^{+0.12}_{-0.11}$ & -1.89$\pm$0.20 & 6.041 & 13.8$\pm$1.5 (10.3$\sigma$) & 20.8$\pm$3.0 \\[2pt]
XMM3-198954 & 02:26:26.29 & -05:08:56.68 & 25.63$^{+0.38}_{-0.28}$ & -3.40$\pm$0.88 & 6.076 & 62.3$\pm$2.7 (23.4$\sigma$) & 107$\pm$32 \\[2pt]
COS-1260899 & 09:58:54.84 & +02:29:12.33 & 25.66$^{+0.16}_{-0.14}$ & -2.43$\pm$0.29 & 6.143 & 22.8$\pm$3.4 (7.0$\sigma$) & 40.7$\pm$8.1 \\[2pt]
COS-631233$^{\ast}$ & 10:00:07.02 & +02:01:48.85 & 26.42$^{+0.34}_{-0.26}$ & -2.98$\pm$0.73 & 6.212 & 31.7$\pm$2.4 (13.2$\sigma$) & 113$\pm$32 \\[2pt]
COS-930465 & 09:58:48.31 & +02:14:33.66 & 25.68$^{+0.15}_{-0.13}$ & -2.57$\pm$0.52 & 6.303 & 9.9$\pm$0.7 (14.1$\sigma$) & 18.4$\pm$2.7 \\[2pt]
\hline
\end{tabular}
\label{tab:LyA_detection_info_z6}
\end{table*}

We also detect confident ($>$7$\sigma$) emission features in 13 of our 30 z$\simeq$6 targets.
In Table \ref{tab:LyA_detection_info_z6}, we report the coordinates, rest-UV slopes, \Lya{} redshifts, fluxes, EWs, and \textit{y}/\textit{Y}-band magnitudes of each detected z$\simeq$6 source in COSMOS/XMM3.
The 2D spectra of all z$\simeq$6 detections are shown in Fig. \ref{fig:LyADetectionMosaic_z6} and span redshifts of \zLya{} = 5.824--6.303, consistent with expectations given their selection criteria (\S\ref{sec:z6_selection}).
We measure \Lya{} fluxes ranging from (6.8--62.3)$\times$10$^{-18}$ erg/s/cm$^2$ and EWs ranging from 5.5--113 \AA{}.

\begin{figure*}
\includegraphics{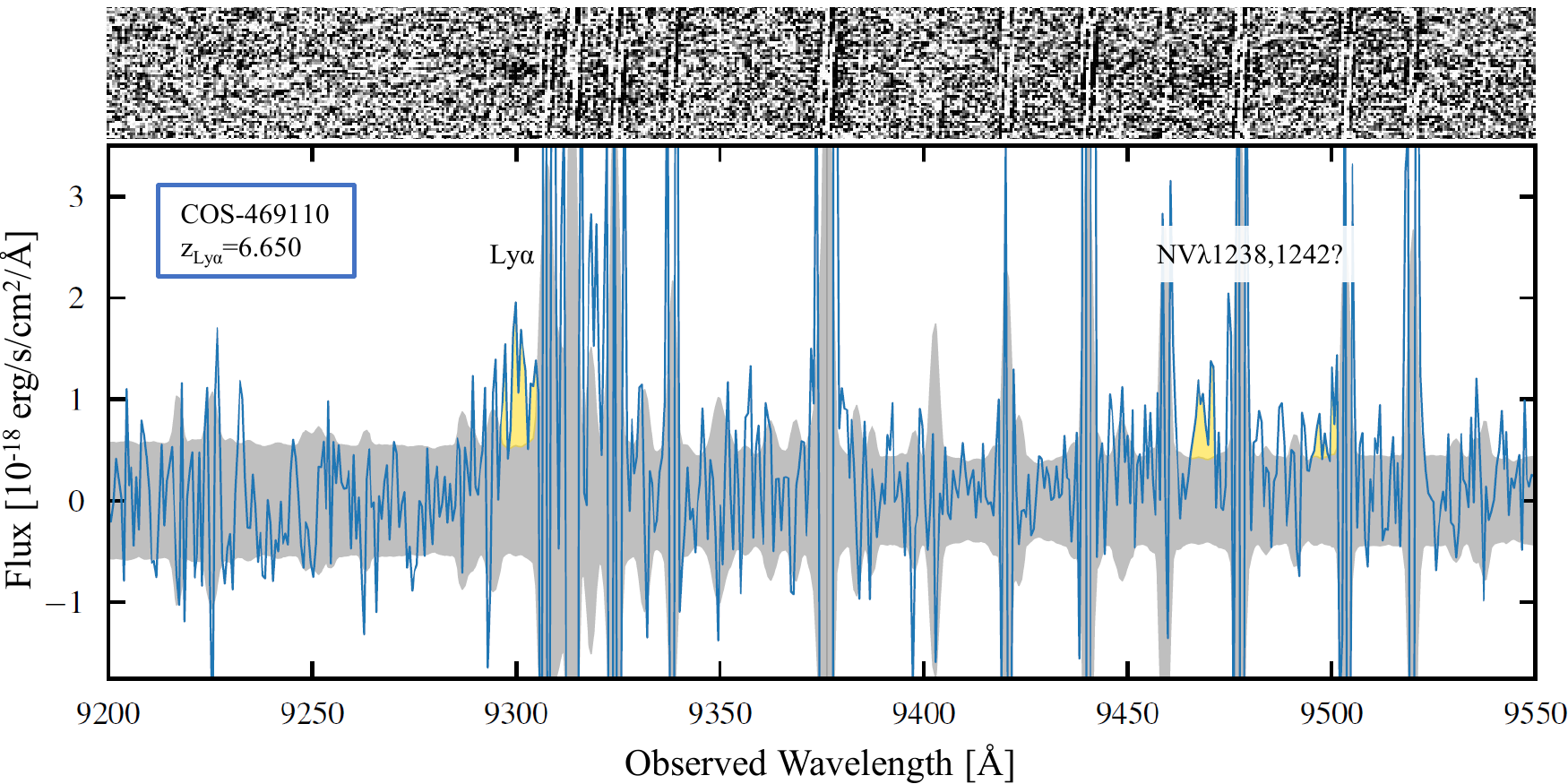}
\caption{MMT/Binospec spectra of COS-469110 where we identify Ly$\alpha$ as well as tentative detections of the NV$\lambda$1238.8,1242.8 doublet, a signpost of significant AGN activity. The top and middle panels show the 2D and smoothed 2D signal-to-noise ratio maps, respectively, where black is positive. In the bottom panel, we show the 1D extraction centred on the expected spatial position of the source with the 1$\sigma$ noise level in gray. The two NV doublet components are measured with a significance of 6.4$\sigma$ and 5.2$\sigma$, respectively, and have peak wavelengths corresponding to the exact same redshift of z=6.645. This translates to a \Lya{} velocity offset of $\approx$200 km s$^{-1}$ which may help explain the exceptionally high NV to \Lya{} flux ratio of this source (see text).}
\label{fig:COS469110_NV}
\end{figure*}

Motivated by recent detections in similarly luminous z$\gtrsim$7 galaxies (\citealt{Tilvi2016,Laporte2017,Mainali2018}; see also \citealt{Hu2017}), we search for NV$\lambda$1238.8,1242.8 emission in the spectra of our z$\simeq$7 \Lya{} emitters.
Given the very high ionization potential of this line (77 eV), any such detection would likely signal significant AGN activity.
As shown in Fig. \ref{fig:COS469110_NV}, we tentatively identify the NV doublet in COS-469110 (\Muv{} = $-$21.7) where the emission is located at the exact expected spatial position of the source.
We regard both these features as tentative because each are detected at $<$7$\sigma$ significance, specifically 6.4$\sigma$ and 5.2$\sigma$ for the 1238.8 and 1242.8 \AA{} components, respectively.
However, we note that if we split our data into two distinct stacks\footnote{We divide our data such that the total weight of all exposures in each stack is roughly equal.}, the NV$\lambda$1238.8 feature is significantly detected in each stack (5.2$\sigma$ and 3.8$\sigma$), adding evidence that this feature is likely real. 

The NV$\lambda$1238.8 component is clearly distinct from strong skylines and has a measured flux of (5.4$\pm$1.0)$\times$10$^{-18}$ erg/s/cm$^2$ which, using the \textit{Y}-band photometry for the continuum, corresponds to an EW of 8.4$\pm$3.0 \AA{}.
The flux of the NV$\lambda$1242.8 component is measured to be (3.8$\pm$0.8)$\times$10$^{-18}$ erg/s/cm$^2$ which corresponds to an EW of 5.9$\pm$2.2 \AA{}.
While the 1242.8 \AA{} component does rest up against a skyline, the peak wavelengths of both components correspond to the exact same redshift of z=6.645 which is well consistent with the \Lya{} redshift of \zLya{} = 6.650 (we discuss the implied \Lya{} velocity offset below). 
Furthermore, the measured EWs of both components are consistent with that recently reported in three other similarly luminous z=7--9 galaxies \citep{Tilvi2016,Laporte2017,Mainali2018} as well as a 1:1 to 2:1 flux ratio for the doublet \citep{Bickel1969,TorresPeimbert1984}.
However, we do note that calculating this flux ratio is complicated by the partial skyline masking of the 1242.8 \AA{} component. 

While we do not detect any significant NV features in any of our other z$\simeq$7 \Lya{} emitters, we are able to place 5$\sigma$ NV$\lambda$1238.8 EW limits\footnote{Here, we report the 5$\sigma$ EW limit over the observed wavelength range corresponding to \Lya{} velocity offsets of 0--500 km s$^{-1}$ similar to the approach of \citet{Mainali2018}.} of $\leq$10 \AA{} in six of our other twelve z$\simeq$7 \Lya{} emitters, suggesting that the (tentative) NV emission from COS-469110 is likely exceptional among the bright z$\simeq$7 population. 
This is perhaps further supported by the fact that the total NV flux we measure implies a line flux ratio of $f_{_{\mathrm{NV}}}/f_{_{\Lya{}}}$ = 0.95$\pm$0.19, much larger than typical upper limits recently placed on several other bright z$>$6.5 \Lya{} emitters ($f_{_{\mathrm{NV}}}/f_{_{\Lya{}}}$ $\lesssim$ 0.2; \citealt{Mainali2018,Shibuya2018}).
Because of the partial skyline obscuration of the the 1242.8 \AA{} component, this flux ratio for COS-469110 is likely a lower limit.

The \Lya{} velocity offset implied for COS-469110 is $\approx$200 km s$^{-1}$, assuming that NV well traces the systemic redshift \citep{Laporte2017}.
This velocity offset falls well within the range previously reported for similarly luminous (\Muv{} $<$ $-$21.5) galaxies at z$>$6 (110--500 km s$^{-1}$; \citealt{Willott2015,Inoue2016,Pentericci2016,Laporte2017,Stark2017,Mainali2018,Matthee2020_VR7}), though it does sit on the lower end of that range.
This may help explain the exceptionally large NV to \Lya{} line flux ratio mentioned above as \Lya{} emission is more susceptible to strong scattering by the partially neutral IGM at lower velocity offsets \citep[e.g.][]{MiraldaEscude1998,Mason2018}.

\begin{figure}
\includegraphics{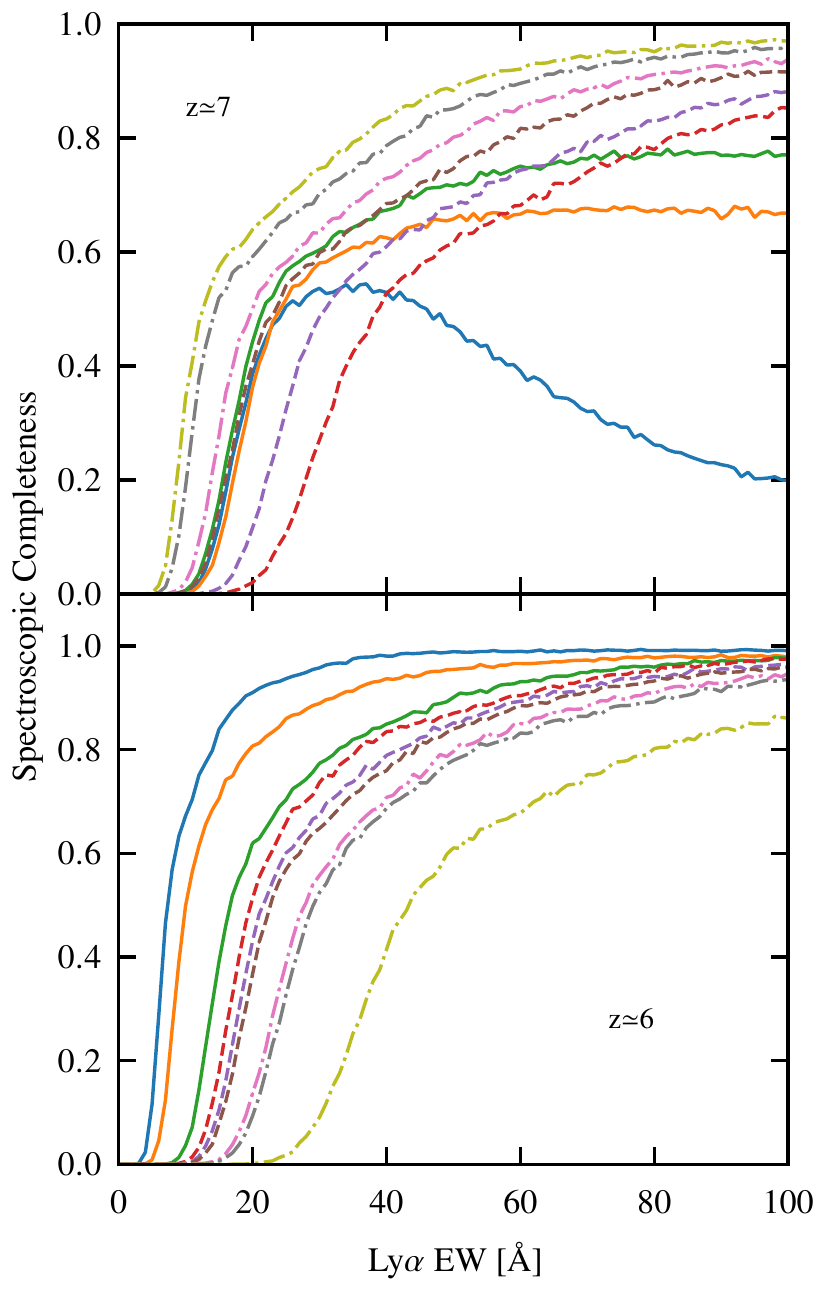}
\caption{Figures showing our simulated spectroscopic completeness as a function of \Lya{} EW for non-detected sources at z$\simeq$7 (top) and z$\simeq$6 (bottom). To improve clarity, we only show the completeness curves for a representative subset of non-detected sources. For a small subset of the non-detected z$\simeq$7 sources, the spectral completeness declines at high \Lya{} EW due to their lack of very red ($\gtrsim$9750 \AA{}) spectral coverage which prohibits detecting \Lya{} at z$\gtrsim$7.0 and such a high-redshift solution is only likely for strong \Lya{} emitters (\S\ref{sec:z7_selection}).
All of this information is folded into our analysis when inferring the \Lya{} EW distributions.}
\label{fig:completeness}
\end{figure}

\subsubsection{Completeness Simulations} \label{sec:completeness}

When inferring the \Lya{} EW distribution in \S\ref{sec:evolution}, we fold in constraints from non-detections. We do so by calculating the \Lya{} detection completeness of each non-detected source as a function of \Lya{} EW from EW = 0--100 \AA{} in 1 \AA{} steps.
We insert 10,000 simulated 1D flux profiles for each value of EW into the fully reduced 1D spectrum of each source at its expected spatial position and determine what fraction of simulated features would be detected at $>$7$\sigma$.
The observed wavelengths of these simulated features are randomly sampled using the redshift completeness distributions as a function of \Lya{} EW and rest-UV magnitude described in \S\ref{sec:sample_selection}.
In doing so, we account for the dependence of our redshift selection window on \Lya{} EW and for the impact of skyline obscuration.
The profile of each simulated \Lya{} feature is a Gaussian with FWHM set to 220 km s$^{-1}$, equal to the mean value measured from the two z$\simeq$7 \Lya{} features detected with very high significance ($>$15$\sigma$) in our sample, namely COS-788571 (FWHM = 180 km s$^{-1}$) and COS-1009842 (FWHM = 260 km s$^{-1}$). 
This FWHM is also in excellent agreement with that found by \citet{Pentericci2018} using stacked spectra of fainter z$\simeq$7 galaxies.

We plot the simulated spectroscopic completeness for a representative subset of the non-detected z$\simeq$7 (z$\simeq$6) galaxies in the top (bottom) panel of Fig. \ref{fig:completeness}. 
The spectral constraints on most of these z$\simeq$6-7 targets yield $\gtrsim$50\% completeness at EW = 20 \AA{}, increasing to $\gtrsim$80\% at EW = 60 \AA{}.
For a small subset of the non-detected z$\simeq$7 sources, the spectral completeness declines at high \Lya{} EW due to their lack of very red ($\gtrsim$9750 \AA{}) spectral coverage which prohibits detecting \Lya{} at z$\gtrsim$7.0.
Such a high-redshift solution is only likely for strong \Lya{} emitters given our selection criteria (\S\ref{sec:z7_selection}).
All of this information is folded into our analysis when inferring the \Lya{} EW distributions (\S\ref{sec:analysis}). 

\begin{table*}
\centering
\caption{Inferred properties of our z$\simeq$7 sample targeted with MMT/Binospec. These properties were obtained by fitting photometry with a photoionization model using the \textsc{beagle} SED fitting code \citep{Chevallard2016}. The best-fitting values and errors are determined by calculating the median and inner 68\% confidence interval values marginalized over the posterior probability distribution function output by \textsc{beagle}. We report spectroscopic redshifts for sources with a confident ($>$7$\sigma$) \Lya{} detection and photometric redshifts otherwise.}
\begin{tabular}{P{1.8cm}P{2.5cm}P{1.1cm}P{1.5cm}P{1.3cm}P{2.0cm}P{2.2cm}P{1.8cm}} 
\hline
Source ID & Redshift & \Muv{} & \logten{} \Mstar{} & $\tau_{_V}$ & sSFR & [OIII]+H$\beta$ EW & \logten{} \xiion{} \Tstrut{} \\
 &  &  & [\Msol{}] &  & [Gyr$^{-1}$] & [\AA{}] & [erg$^{-1}$ Hz] \Bstrut{} \\
\hline
COS-221419 & 6.70$^{+0.05}_{-0.04}$ & -21.0$^{+0.1}_{-0.1}$ & 8.4$^{+0.4}_{-0.4}$ & 0.01$^{+0.02}_{-0.01}$ & 6.3$^{+47.3}_{-5.8}$ & 690$^{+760}_{-440}$ & 25.52$^{+0.19}_{-0.25}$ \Tstrut{} \\[4pt]
COS-235129 & 6.76$^{+0.08}_{-0.07}$ & -21.4$^{+0.1}_{-0.1}$ & 9.5$^{+0.3}_{-1.0}$ & 0.03$^{+0.08}_{-0.02}$ & 0.9$^{+4.2}_{-0.7}$ & 200$^{+230}_{-150}$ & 25.25$^{+0.32}_{-0.39}$ \\[4pt]
COS-237729 & 6.83$^{+0.10}_{-0.09}$ & -21.1$^{+0.1}_{-0.1}$ & 8.8$^{+0.8}_{-0.7}$ & 0.05$^{+0.09}_{-0.05}$ & 5.1$^{+37.1}_{-4.5}$ & 650$^{+530}_{-420}$ & 25.60$^{+0.21}_{-0.29}$ \\[4pt]
COS-301652 & 6.63$^{+0.09}_{-0.03}$ & -21.2$^{+0.1}_{-0.1}$ & 9.3$^{+0.7}_{-1.1}$ & 0.08$^{+0.10}_{-0.07}$ & 2.0$^{+9.5}_{-1.5}$ & 520$^{+330}_{-270}$ & 25.59$^{+0.18}_{-0.25}$ \\[4pt]
COS-469110 & 6.650 & -21.7$^{+0.2}_{-0.1}$ & 8.8$^{+0.8}_{-0.5}$ & 0.04$^{+0.09}_{-0.04}$ & 5.3$^{+32.4}_{-4.7}$ & 730$^{+380}_{-320}$ & 25.58$^{+0.22}_{-0.20}$ \\[4pt]
COS-505871 & 6.67$^{+0.12}_{-0.07}$ & -21.2$^{+0.1}_{-0.1}$ & 10.0$^{+0.2}_{-1.5}$ & 0.04$^{+0.15}_{-0.03}$ & 1.2$^{+3.4}_{-0.7}$ & 290$^{+320}_{-190}$ & 25.50$^{+0.24}_{-0.31}$ \\[4pt]
COS-534584 & 6.60$^{+0.01}_{-0.01}$ & -21.9$^{+0.1}_{-0.1}$ & 9.2$^{+0.5}_{-0.7}$ & 0.12$^{+0.04}_{-0.06}$ & 5.7$^{+26.2}_{-5.1}$ & 790$^{+520}_{-390}$ & 25.50$^{+0.20}_{-0.19}$ \\[4pt]
COS-788571 & 6.884 & -21.5$^{+0.1}_{-0.1}$ & 8.9$^{+0.4}_{-0.3}$ & 0.02$^{+0.06}_{-0.01}$ & 173.3$^{+253.5}_{-124.7}$ & 3680$^{+1940}_{-1660}$ & 25.91$^{+0.06}_{-0.11}$ \\[4pt]
COS-851423 & 6.65$^{+0.05}_{-0.03}$ & -21.1$^{+0.1}_{-0.1}$ & 8.3$^{+0.7}_{-0.3}$ & 0.01$^{+0.03}_{-0.01}$ & 9.5$^{+55.8}_{-8.9}$ & 1030$^{+730}_{-480}$ & 25.63$^{+0.14}_{-0.19}$ \\[4pt]
COS-854905 & 6.67$^{+0.05}_{-0.03}$ & -21.4$^{+0.1}_{-0.1}$ & 8.8$^{+0.9}_{-0.6}$ & 0.02$^{+0.05}_{-0.01}$ & 2.4$^{+14.5}_{-2.0}$ & 560$^{+490}_{-350}$ & 25.51$^{+0.20}_{-0.28}$ \\[4pt]
COS-856875 & 6.68$^{+0.05}_{-0.03}$ & -21.2$^{+0.1}_{-0.1}$ & 8.6$^{+0.6}_{-0.5}$ & 0.01$^{+0.03}_{-0.01}$ & 3.1$^{+25.0}_{-2.7}$ & 540$^{+430}_{-280}$ & 25.48$^{+0.20}_{-0.22}$ \\[4pt]
COS-862541 & 6.850 & -22.5$^{+0.1}_{-0.1}$ & 9.1$^{+0.4}_{-0.3}$ & 0.06$^{+0.07}_{-0.05}$ & 155.6$^{+193.2}_{-123.2}$ & 4160$^{+1610}_{-1270}$ & 25.89$^{+0.08}_{-0.08}$ \\[4pt]
COS-940214 & 6.748 & -20.4$^{+0.2}_{-0.2}$ & 8.0$^{+0.4}_{-0.3}$ & 0.01$^{+0.04}_{-0.01}$ & 134.1$^{+238.5}_{-122.8}$ & 3210$^{+1950}_{-1790}$ & 25.87$^{+0.10}_{-0.16}$ \\[4pt]
COS-955126 & 6.813 & -21.5$^{+0.1}_{-0.1}$ & 8.5$^{+0.6}_{-0.3}$ & 0.02$^{+0.06}_{-0.02}$ & 29.2$^{+97.5}_{-27.6}$ & 1620$^{+980}_{-650}$ & 25.73$^{+0.10}_{-0.11}$ \\[4pt]
COS-1009842 & 6.761 & -20.6$^{+0.2}_{-0.1}$ & 8.3$^{+0.8}_{-0.4}$ & 0.03$^{+0.07}_{-0.02}$ & 10.4$^{+58.1}_{-9.3}$ & 910$^{+720}_{-500}$ & 25.65$^{+0.15}_{-0.22}$ \\[4pt]
COS-1048848 & 6.69$^{+0.11}_{-0.08}$ & -20.7$^{+0.1}_{-0.1}$ & 8.4$^{+0.3}_{-0.3}$ & 0.01$^{+0.03}_{-0.01}$ & 2.1$^{+15.8}_{-1.8}$ & 310$^{+300}_{-180}$ & 25.35$^{+0.24}_{-0.33}$ \\[4pt]
COS-1053257 & 6.68$^{+0.07}_{-0.05}$ & -22.0$^{+0.1}_{-0.1}$ & 8.5$^{+0.8}_{-0.2}$ & 0.04$^{+0.06}_{-0.04}$ & 5.0$^{+42.3}_{-4.6}$ & 630$^{+530}_{-310}$ & 25.75$^{+0.13}_{-0.23}$ \\[4pt]
COS-1099982 & 6.68$^{+0.03}_{-0.02}$ & -21.6$^{+0.1}_{-0.1}$ & 8.6$^{+0.5}_{-0.3}$ & 0.01$^{+0.03}_{-0.01}$ & 72.7$^{+112.0}_{-54.9}$ & 2470$^{+1080}_{-690}$ & 25.76$^{+0.09}_{-0.07}$ \\[4pt]
COS-1205190 & 7.049 & -20.9$^{+0.2}_{-0.2}$ & 8.6$^{+0.4}_{-0.5}$ & 0.01$^{+0.03}_{-0.01}$ & 2.8$^{+25.4}_{-2.4}$ & 330$^{+470}_{-220}$ & 25.42$^{+0.27}_{-0.40}$ \\[4pt]
COS-1235751 & 6.74$^{+0.09}_{-0.07}$ & -21.3$^{+0.1}_{-0.1}$ & 9.4$^{+0.5}_{-0.6}$ & 0.32$^{+0.06}_{-0.07}$ & 1.6$^{+9.9}_{-1.3}$ & 300$^{+240}_{-190}$ & 25.35$^{+0.30}_{-0.34}$ \\[4pt]
XMM3-227436 & 7.093 & -22.3$^{+0.2}_{-0.1}$ & 9.0$^{+0.7}_{-0.4}$ & 0.04$^{+0.09}_{-0.03}$ & 6.6$^{+56.5}_{-6.0}$ & 930$^{+800}_{-510}$ & 25.84$^{+0.28}_{-0.23}$ \\[4pt]
XMM3-504799 & 6.883 & -22.5$^{+0.1}_{-0.1}$ & 9.3$^{+0.6}_{-0.4}$ & 0.05$^{+0.10}_{-0.04}$ & 80.1$^{+190.9}_{-71.3}$ & 2310$^{+1830}_{-1150}$ & 26.22$^{+0.25}_{-0.23}$ \\[4pt]
\hline
\end{tabular}
\label{tab:galaxy_properties}
\end{table*}

\begin{figure*}
\includegraphics{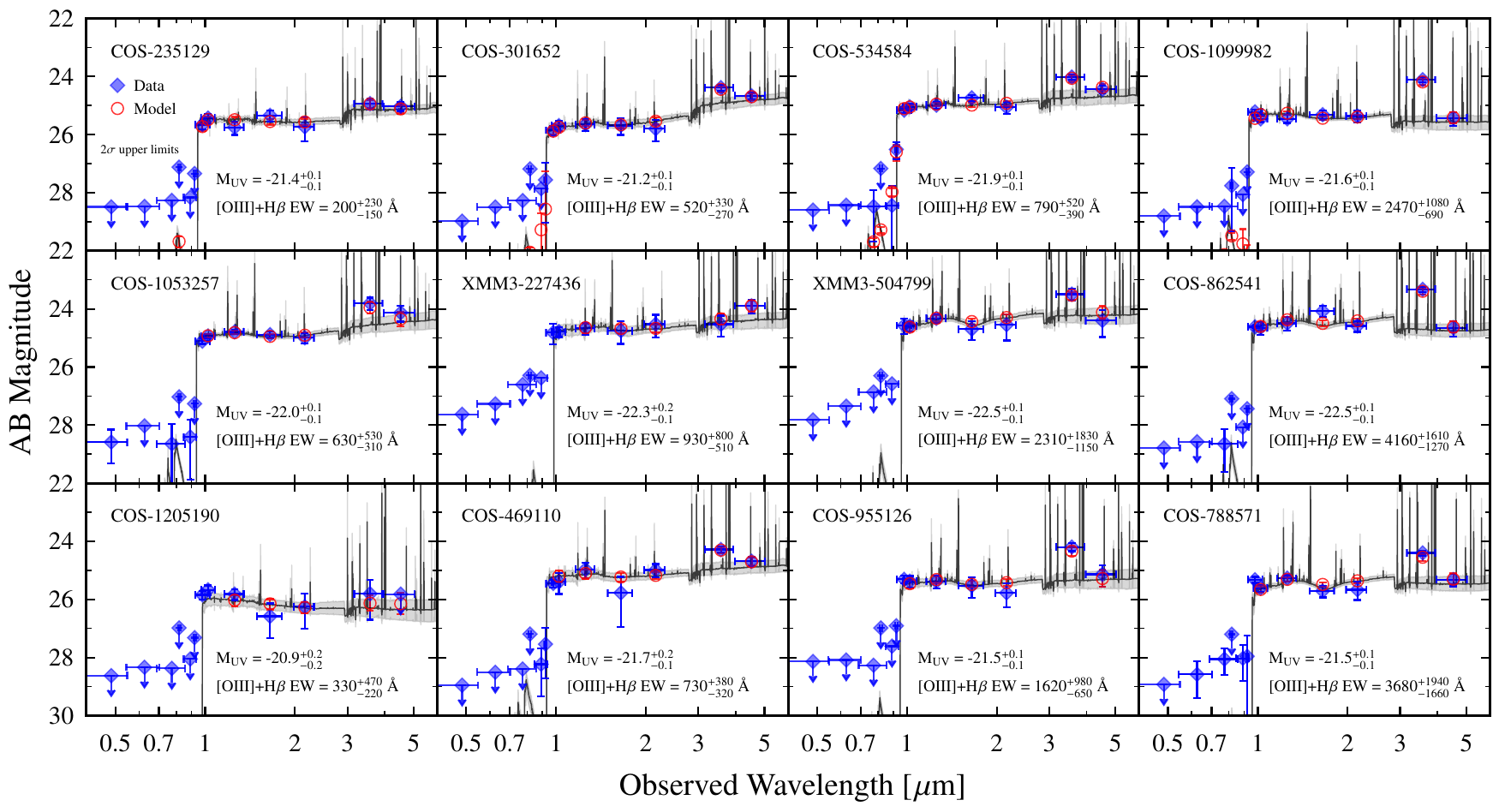}
\caption{Spectral energy distributions of a subset of z$\simeq$7 sources in our spectroscopic sample. We select sources to display the variety of absolute UV magnitudes, \Muv{}, and inferred [OIII]$+$H$\beta$ EWs (generally increasing to the right) in this sample. In each panel, the median fit \textsc{beagle} SED model is overlaid in black with gray shading showing the inner 68\% confidence interval from the posterior output by \textsc{beagle}. Blue diamonds show the fitted photometry (with 2$\sigma$ upper limits in cases of non-detections) while the red empty circles show the median model photometry from \textsc{beagle} with errors enclosing the 68\% confidence interval. We only show model photometry for bands used in the fitting process where we ignore bands impacted by \Lya{} emission or bluewards of the \Lya{} break for sources with a $>$7$\sigma$ \Lya{} detection.}
\label{fig:variousSEDs}
\end{figure*}

\subsection{Photoionization Modeling} \label{sec:BEAGLE}

We now infer the \OIIIHb{} EWs of each of our z$\simeq$7 targets to test whether we see a correlation with \Lya{} visibility.
To infer the \OIIIHb{} emission strength (as well as other physical properties such as  stellar mass), we use the BayEsian Analysis of GaLaxy sEds (\textsc{beagle}; \citealt{Chevallard2016}) SED fitting code.
\textsc{beagle} computes both the stellar and nebular emission of star-forming galaxies by adopting the photoionization models from \citet{Gutkin2016}, which are derived by incorporating the latest version of \citet{BruzualCharlot2003} stellar population synthesis models into \textsc{cloudy} \citep{Ferland2013}. 
Our SED fitting procedure with \textsc{beagle} matches that described in \citetalias{Endsley2021}. 
Briefly, we assume a delayed star formation history (SFR $\propto$ t e$^{-t/\tau}$) with an allowed recent ($<$10 Myr) burst, a minimum age of 1 Myr, an sSFR ranging from 0.1 Gyr$^{-1}$ to 1000 Gyr$^{-1}$, and an SMC dust prescription. 

We note that some of our spectroscopically confirmed sources lie at \zLya{} $\approx$ 6.85--6.9 (COS-788571, COS-862541, and XMM3-504799) where the transmission of [OIII]$\lambda$5007 through [3.6] (and hence the inferred \OIIIHb{} EW) is very sensitive to the exact systemic redshift, z$_{\mathrm{sys}}$.
At z$\gtrsim$6, visible \Lya{} emission is often redshifted relative to systemic due to complex radiative transfer effects in the ISM, CGM, and IGM (see e.g. \citealt{Dijkstra2014} for a review).
The current maximum observed velocity offset at z$>$6 is 500 km s$^{-1}$ \citep{Willott2015} which translates to z$_{\mathrm{sys}}$ = \zLya{}$-$0.013 at z=6.6--7.1 where we have z$\simeq$7 \Lya{} detections.
Therefore, during the \textsc{beagle} fitting process, we allow z$_{\mathrm{sys}}$ to range between \zLya{}$-$0.013 and \zLya{} for z$\simeq$7 sources with a \Lya{} detection.

For sources with a \Lya{} detection, we furthermore exclude bands blueward of the \Lya{} break during the fitting process.
We also do not fit to bands significantly impacted by \Lya{} emission for these sources as it is currently not possible to leave the effective \Lya{} transmission fraction through the IGM as a free parameter in \textsc{beagle}.
This means we fit to all bands redder than and including \textit{Y} for sources with \zLya{}$<$6.9 and all bands redder than including \textit{J} for sources with \zLya{}$>$6.9.
Sources without a \Lya{} detection are fit using all optical through mid-infrared photometry with a uniform redshift prior of z=6--8 and we remove \Lya{} emission from the nebular templates given the low ($\sim$10\%) fraction of strong \Lya{} emitters (EW$>$25 \AA{}) found among the bright z$\simeq$7 population in previous works \citep[e.g.][]{Ono2012,Schenker2014,Pentericci2018}.
We report the inferred [OIII]$+$H$\beta$ EWs of each z$\simeq$7 source in Table \ref{tab:galaxy_properties}, along with their inferred absolute UV magnitudes (at 1600 \AA{} rest-frame), stellar masses, V-band optical depths, and sSFRs.
In the final column, we report the inferred ionizing photon production efficiencies, \xiion{}, computed using the intrinsic UV luminosity (at 1500 \AA{} rest-frame) of the stellar population before processing through dust and gas (see \citealt{Chevallard2018_z0} for details).
We also quote photometric redshifts for sources without a \Lya{} detection and \zLya{} for sources with a \Lya{} detection in Table \ref{tab:galaxy_properties}.

Overall, we find a diverse range of galaxy properties within our sample as illustrated in Fig. \ref{fig:variousSEDs}.
In particular, the inferred [OIII]$+$H$\beta$ EWs range from 200--4000 \AA{} and the sSFRs span $\sim$1--150 Gyr$^{-1}$.
Galaxies with inferred \OIIIHb{} EW$>$800 \AA{} (and therefore similar to those in the \citetalias{RobertsBorsani2016} sample) tend to possess very large sSFRs ($\gtrsim$30 Gyr$^{-1}$), consistent with expectations of galaxies experiencing a recent strong upturn or burst in star formation activity (\citealt{Atek2011}; \citealt{Amorin2015}; \citealt{Maseda2018}; \citealt{Tang2019}; \citetalias{Endsley2021}). 
As expected \citep{Chevallard2018_z0,Tang2019}, we also find that \xiion{} tends to increase with \OIIIHb{} EW among our sample where sources with \OIIIHb{} EW$>$800 \AA{} are inferred to produce ionizing photons 2.2$\times$ as efficiently as those with weaker \OIIIHb{}.
The median inferred \OIIIHb{} EW of all sources is 710 \AA{}, suggesting that our sample is reasonably well representative of the global z$\simeq$7 galaxy population (typical EW = 692 \AA{}; \citetalias{Endsley2021}).

The absolute UV magnitudes of our z$\simeq$7 targets also span $-$22.5 $\leq$ \Muv{} $\leq$ $-$20.4 which corresponds to a luminosity range of 0.8--5.8 L$_{\mathrm{UV}}^{\ast}$ when adopting the z$\simeq$7 $\mathrm{M_{UV}^{\ast}}$ value\footnote{As appropriate for our work, we are using the (double-power law) luminosity function parameters obtained when treating ground-based as a single galaxy.} of $-$20.6 from \citet{Bowler2017}.
Our survey is therefore capturing a subset of the most luminous reionization-era galaxies.
Furthermore, over half the galaxies in our z$\simeq$7 sample are $>$2 L$_{\mathrm{UV}}^{\ast}$, a population that has been largely missed by previous spectroscopy targeting this epoch.
Considering the published literature of z$\simeq$7 \Lya{} observations from \citet{Fontana2010}, \citet{Pentericci2011,Pentericci2014}, \citet{Vanzella2011}, and \citet{Schenker2012,Schenker2014}, only 4 of 64 targeted galaxies were $>$2 L$_{\mathrm{UV}}^{\ast}$.
The only previous spectroscopic z$\simeq$7 studies with a comparable luminosity range to our sample are \citet{Ono2012} and \citet{Furusawa2016}. 
Even so, our sample is at least twice as large as either of these previous studies.

The bright UV luminosities of our z$\simeq$7 targets suggest that they are among the most massive galaxies present in the reionization era, given both clustering measurements \citep{BaroneNugent2014} as well as inferences of the \Muv{}-\Mstar{} relation \citep{Song2016}. 
Stellar masses derived from our SED fitting procedure suggest a similar picture.
Most of our z$\simeq$7 galaxies are inferred to have \logMstar{}$\sim$9--10, which is significantly more massive than fainter z$\simeq$7 galaxies identified over the GOODS fields (\citetalias{Endsley2021}).
Therefore, in what follows, we interchangeably use the terms `bright' and `massive' when describing our sample.

\begin{table}
\centering
\caption{Compiled information on our z$\simeq$7 sample connecting their physical properties (i.e. [OIII]$+$H$\beta$ EWs and rest-UV slopes) to their \Lya{} EWs. For sources without a \Lya{} detection, we quote the 7$\sigma$ limiting EW in skyline-free regions.}
\begin{tabular}{P{1.8cm}P{1.2cm}P{1.8cm}P{1.5cm}} 
\hline
Source ID & \Lya{} EW & [OIII]+H$\beta$ EW & $\beta$ \Tstrut{} \\
 & [\AA{}] & [\AA{}] & \Bstrut{} \\
\hline 
COS-221419 & $<$11.8 & 690$^{+760}_{-440}$ & $-$2.80$\pm$0.62 \Tstrut{}\\[4pt]
COS-235129 & $<$7.6 & 200$^{+230}_{-150}$ & $-$2.09$\pm$0.47 \\[4pt]
COS-237729 & $<$12.9 & 650$^{+530}_{-420}$ & $-$1.92$\pm$0.34 \\[4pt]
COS-301652 & $<$9.8 & 520$^{+330}_{-270}$ & $-$2.08$\pm$0.09 \\[4pt]
COS-469110 & 12.5$\pm$4.3 & 730$^{+380}_{-320}$ & $-$1.69$\pm$0.50 \\[4pt]
COS-505871 & $<$11.5 & 290$^{+320}_{-190}$ & $-$2.23$\pm$0.62 \\[4pt]
COS-534584 & $<$5.2 & 790$^{+520}_{-390}$ & $-$1.76$\pm$0.29 \\[4pt]
COS-788571 & 30.6$\pm$3.9 & 3680$^{+1940}_{-1660}$ & $-$2.11$\pm$0.53 \\[4pt]
COS-851423 & $<$7.5 & 1030$^{+730}_{-480}$ & $-$2.64$\pm$0.38 \\[4pt]
COS-854905 & $<$6.4 & 560$^{+490}_{-350}$ & $-$1.95$\pm$0.24 \\[4pt]
COS-856875 & $<$20.0 & 540$^{+430}_{-280}$ & $-$2.07$\pm$0.37 \\[4pt]
COS-862541 & 11.8$\pm$2.7 & 4160$^{+1610}_{-1270}$ & $-$1.90$\pm$0.43 \\[4pt]
COS-940214 & 43.1$\pm$14.7 & 3210$^{+1950}_{-1790}$ & $-$2.77$\pm$0.54 \\[4pt]
COS-955126 & 12.3$\pm$2.5 & 1620$^{+980}_{-650}$ & $-$2.44$\pm$0.13 \\[4pt]
COS-1009842 & 41.6$\pm$9.5 & 910$^{+720}_{-500}$ & $-$2.61$\pm$0.44 \\[4pt]
COS-1048848 & $<$11.2 & 310$^{+300}_{-180}$ & $-$2.43$\pm$0.35 \\[4pt]
COS-1053257 & $<$3.2 & 630$^{+530}_{-310}$ & $-$2.02$\pm$0.20 \\[4pt]
COS-1099982 & $<$12.5 & 2470$^{+1080}_{-690}$ & $-$1.83$\pm$0.08 \\[4pt]
COS-1205190 & 28.8$\pm$6.0 & 330$^{+470}_{-220}$ & $-$3.44$\pm$1.03 \\[4pt]
COS-1235751 & $<$17.4 & 300$^{+240}_{-190}$ & $-$1.10$\pm$0.30 \\[4pt]
XMM3-227436 & 15.0$\pm$3.2 & 930$^{+800}_{-510}$ & $-$1.85$\pm$0.27 \\[4pt]
XMM3-504799 & 3.7$\pm$0.8 & 2310$^{+1830}_{-1150}$ & $-$2.02$\pm$0.42 \\[4pt]
\hline
\end{tabular}
\label{tab:shortTable}
\end{table}

\section{Analysis} \label{sec:analysis}

In this section, we first investigate how \Lya{} detectability correlates with \OIIIHb{} emission (and hence sSFR) within our sample of 22 bright (L$_{\mathrm{UV}}^{}$ $\simeq$ 1--6 L$_{\mathrm{UV}}^{\ast}$) z$\simeq$7 galaxies (\S\ref{sec:z7_LyA}).
We then quantify evolution in the \Lya{} EW distribution of bright galaxies between z$\simeq$6--7 to test whether our results are consistent with accelerated reionization around massive z$\simeq$7 systems (\S\ref{sec:evolution}). 
Finally, we investigate the spatial separations of our z$\simeq$7 \Lya{} emitters to identify any potentially large ionized bubbles within our observed fields (\S\ref{sec:bubbles}).

\begin{figure}
\includegraphics{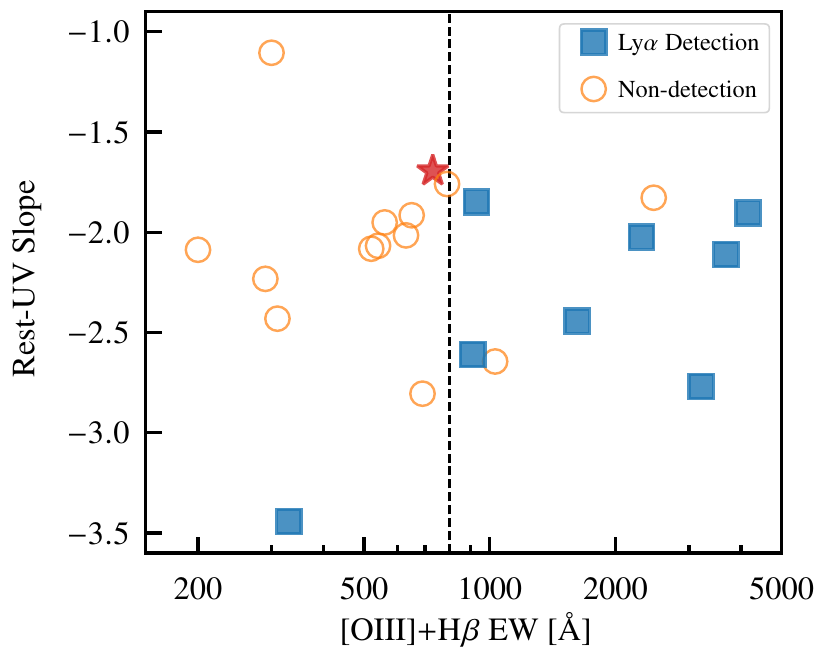}
\caption{Plot showing the inferred \OIIIHb{} EWs and measured rest-UV slopes for each of our 22 massive (L$_{\mathrm{UV}}^{}$ $\simeq$ 1--6 L$_{\mathrm{UV}}^{\ast}$) z$\simeq$7 systems. We mark those with and without confident ($>$7$\sigma$) \Lya{} detections as shown in the legend. As indicated by the prominence of blue markers to the right of the plot, we confidently detect \Lya{} at a much higher rate from strong \OIIIHb{} emitters ($>$800 \AA{} EW; 78\% detection rate) than more moderate \OIIIHb{} emitters (200-800 \AA{} EW; 8\% detection rate). Our results therefore suggest that \Lya{} is more readily detectable from massive z$\simeq$7 galaxies experiencing a rapid upturn in star formation activity (i.e. high sSFR). We also use a red star to show COS-469110 which likely harbors an AGN given its tentative NV detection (see Fig. \ref{fig:COS469110_NV}).}
\label{fig:whichDetected}
\end{figure}

\begin{figure}
\includegraphics{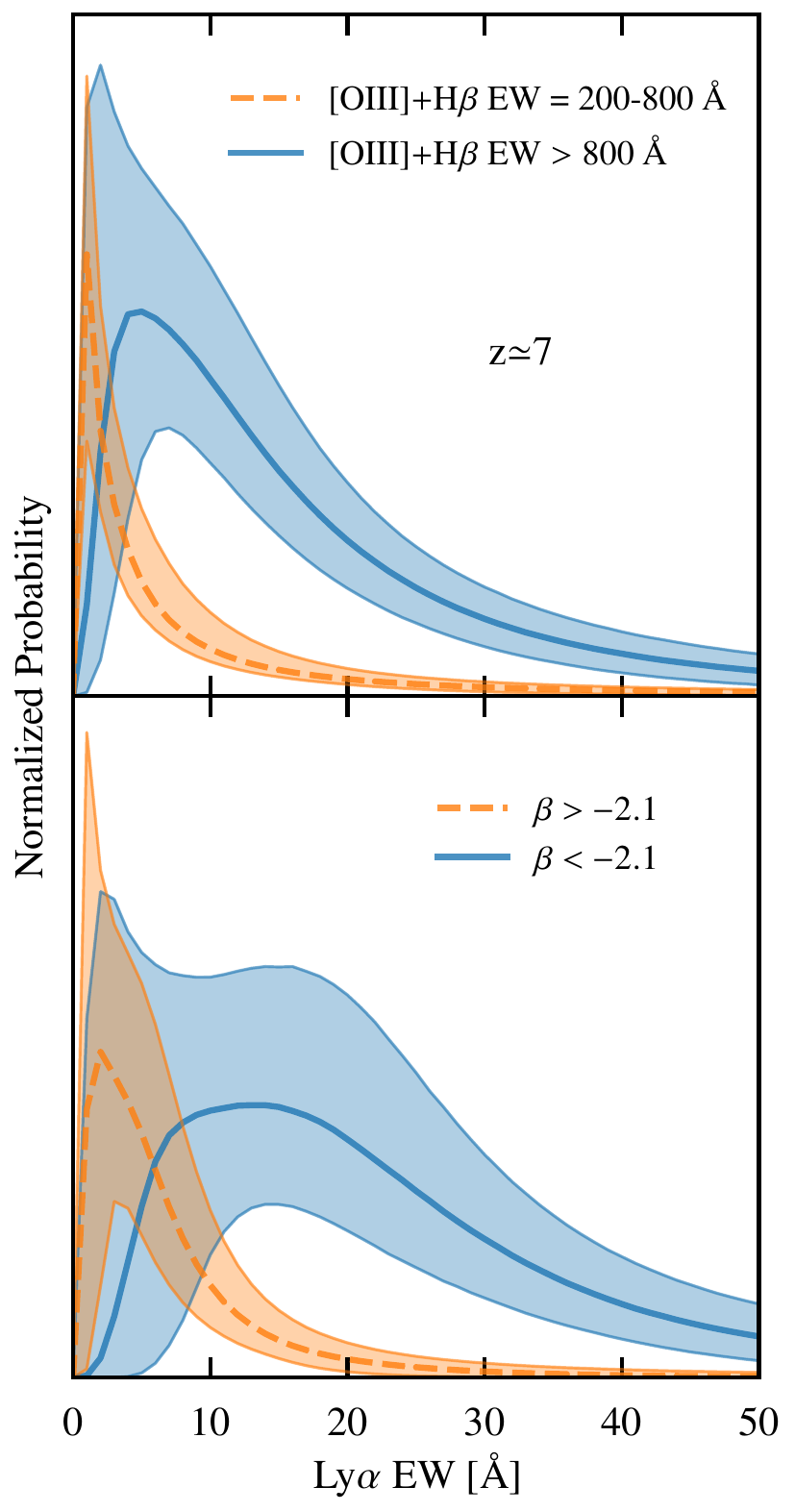}
\caption{Comparison of the inferred \Lya{} EW distributions for our z$\simeq$7 sample as a function of physical properties. In the top panel, we compare the \Lya{} EW distributions for moderate (orange dashed curve) versus strong \OIIIHb{} emitters (solid blue). In the bottom panel, we compare the inferred \Lya{} EW distributions for our z$\simeq$7 galaxies with rest-UV slopes of $\beta > -2.1$ (dashed orange) versus the bluest sources with $\beta < -2.1$ (solid blue curve). These plots demonstrate that massive z$\simeq$7 galaxies with strong \OIIIHb{} emission and/or very blue rest-UV slopes show evidence for systematically higher EW \Lya{} emission.}
\label{fig:z7EWDistnComparison}
\end{figure}

\subsection{The Connection Between [OIII]$+$H$\beta$ and Ly$\alpha$ at z$\simeq$7} \label{sec:z7_LyA}

The number of \Lya{} detections among UV-continuum selected galaxies in the reionization era (z$\gtrsim$7) has grown rapidly over the past decade \citep[e.g.][]{Vanzella2011,Ono2012,Schenker2012,Shibuya2012,Finkelstein2013,Pentericci2014,Pentericci2018,Song2016_LyA,Jung2017,Jung2018,Castellano2018,Larson2018,Hoag2019,Fuller2020,Tilvi2020}.
One of the most surprising results from these campaigns was the 100\% \Lya{} detection rate among the four brightest (L$_{\mathrm{UV}}^{}$ = 3--4 L$_{\mathrm{UV}}^{\ast}$) z$>$7 galaxies selected over the CANDELS fields (i.e. the \citetalias{RobertsBorsani2016} sample).
While this led some to suggest that massive reionization-era systems may trace accelerated sites of reionization \citep{Zitrin2015,Stark2017}, this interpretation was complicated by the fact that the \citetalias{RobertsBorsani2016} galaxies were selected to not only be very massive, but also to possess strong \OIIIHb{} emission (EW$>$800 \AA{}). 
Such strong nebular emission signals the presence of intense radiation fields likely powered by a recent rapid upturn in star formation activity (i.e. high sSFR; \citealt{Tang2019}; \citetalias{Endsley2021}).
This therefore raises the possibility that the \Lya{} detections within the \citetalias{RobertsBorsani2016} sample were not necessarily due to large ionized regions, but perhaps instead driven by physics internal to each of the four galaxies \citep{Tang2019}.

To better understand the origin of strong \Lya{} from the \citetalias{RobertsBorsani2016} sample, we here analyse the \Lya{} constraints from our larger sample (N=22) of similarly massive (L$_{\mathrm{UV}}^{}$ $\simeq$ 1--6 L$_{\mathrm{UV}}^{\ast}$) z$\simeq$7 galaxies possessing a wide range of inferred \OIIIHb{} EWs (200--4000 \AA{}).
This diversity of nebular emission strengths enables us to explore how the sSFRs of massive reionization-era galaxies impact their \Lya{} detectability.
To this end, we divide our sample into galaxies with strong (EW$>$800 \AA{}) versus more moderate (EW = 200--800 \AA{}) inferred \OIIIHb{} emission.
By this classification, galaxies in the strong \OIIIHb{} emitter sub-sample will possess large sSFRs comparable to those in the \citetalias{RobertsBorsani2016} sample while the moderate \OIIIHb{} emitters are more representative of the typical massive z$\simeq$7 population (\citetalias{Endsley2021}).
Here, we do not include COS-469110 in either sub-sample given its tentative NV detection (Fig. \ref{fig:COS469110_NV}) which is a likely signpost of significant AGN activity.
We do, however, discuss the potential implications of this unique source below.

An essential question for interpreting the \citetalias{RobertsBorsani2016} result is whether our z$\simeq$7 targets with strong \OIIIHb{} emission show an increased \Lya{} detection rate.
From our observations, we confidently ($>$7$\sigma$) detect \Lya{} in seven of nine (78\%) galaxies with strong \OIIIHb{} emission, yet in only one of twelve (8\%) galaxies with more moderate \OIIIHb{} (see Fig. \ref{fig:whichDetected}).
Our results therefore suggest that \Lya{} is indeed more easily detectable from massive z$\simeq$7 galaxies possessing very large sSFRs ($\gtrsim$30 Gyr$^{-1}$). 
We note that the two strong \OIIIHb{} emitters in our sample that went undetected may exhibit prominent \Lya{} emission that is masked by skylines.
This possibility is accounted for in our analysis below.

Given this marked contrast in \Lya{} detection rate, we now seek to quantify the enhancement in \Lya{} EW among those with strong \OIIIHb{} relative to the more moderate population.
To do so, we infer the \Lya{} EW distribution of each \OIIIHb{} emitter sub-sample utilizing a Bayesian approach that accounts for spectroscopic incompleteness of non-detected sources.
Specifically, we assume that the \Lya{} EW distribution of each sub-sample follows a log-normal\footnote{We adopt a log-normal EW distribution throughout our analysis because this function yields the best fit to our data when comparing to a Gaussian \citep[e.g.][]{Ouchi2008,Guaita2010} and exponentially declining \citep[e.g.][]{Jung2017,Mason2018_IGMneutralFrac} function. None of our major conclusions are significantly altered if we adopt one of these other distributions instead.} function \citep[e.g.][]{Schenker2014} parametrized by a median EW, \muEW{}, and standard deviation, \sigmaEW{}.
After generating a grid\footnote{Here we ignore the third parameter, $A$, in the parametrization by \citet{Schenker2014} used to quantify the fraction of sources with \Lya{} EW$>$0. We find that this value is well consistent with unity when constraining the EW distribution using all 22 z$\simeq$7 sources in our sample (\S\ref{sec:evolution}).} spanning \logten{}(\muEW{}/\AA{}) = 0.0--2.0 \AA{} and \sigmaEW{} = 0.1--1.2 dex (both with a spacing of 0.01), we compute the probability for each set of parameters, $P(\muEW{},\sigmaEW{})$, following the Bayesian approach of Boyett et al. (2020 in prep; see also \citetalias{Endsley2021}):
\begin{equation} \label{eq:prob}
P \left(\muEW{},\sigmaEW{}\right) \propto \prod_i \int P_i\left(\mathrm{EW}\right) P\left(EW\ |\ \muEW{},\sigmaEW{}\right)\  d\mathrm{EW}.
\end{equation}
Here, $i$ represents the index of each z$\simeq$7 source and, for sources with a \Lya{} detection, $P_i(\mathrm{EW})$ is a Gaussian distribution centred on the measured EW with its calculated uncertainties.
For those without a \Lya{} detection, $P_i(\mathrm{EW})$ is the probability that a \Lya{} line with equivalent width EW would not have been detected.
This is set equal to $1-\mathscr{C}_i(\mathrm{EW})$ where $\mathscr{C}_i(\mathrm{EW})$ is the spectroscopic completeness for source $i$ calculated using the simulations described in \S\ref{sec:completeness} (see also Fig. \ref{fig:completeness}).

We find that massive z$\simeq$7 galaxies with strong \OIIIHb{} emission typically exhibit much stronger \Lya{} (4.7$\times$ higher EW) relative to more moderate \OIIIHb{} emitters when comparing the best-fitting \muEW{} values for each sub-sample (see top panel of Fig. \ref{fig:z7EWDistnComparison}). 
Therefore, while uncertainties remain significant\footnote{The median and inner 68\% confidence interval of this ratio on median EWs from the marginalized $P(\muEW{})$ of each sub-sample is 3.8$^{\scaleto{+4.4}{4.5pt}}_{\scaleto{-2.4}{4.5pt}}$. These uncertainties are largely due to our current sample size and non-detections in most moderate \OIIIHb{} emitters.}, our results suggest that galaxy sSFRs play a significant role in regulating \Lya{} emission from massive reionization-era galaxies.
The recent results of \citet{Castellano2017} provide additional empirical support of this picture.
In their study, they find that the stacked IRAC colours of z$\simeq$6.8 \Lya{} emitters selected over the CANDELS fields indicate much stronger \OIIIHb{} emission relative to galaxies that went undetected in \Lya{}.
This is consistent with our results using \OIIIHb{} EWs inferred for individual (and generally more massive) galaxies.

Interestingly, two sources in our sample, COS-469100 and COS-1205190, both show fairly strong \Lya{} (EW=15--29 \AA{}), yet possess only moderate \OIIIHb{} emission (EW=200-800 \AA{}).
It is therefore of interest to explore whether there are other reasons to believe that these two sources may still possess unusually powerful radiation fields as expected in very large sSFR systems.
Indeed, the tentative NV detection within COS-469110 (see Fig. \ref{fig:COS469110_NV}) signals that this source likely harbors an AGN.
We also find reason to believe that COS-1205190 may be powering an intense radiation field given its extremely blue rest-UV slope of $\beta$ = $-$3.44$\pm$1.03.
Such a blue slope (the bluest in our sample; Table \ref{tab:shortTable}) is consistent with not only extremely low dust content, but also very low metallicity.
This possible dearth of metals in COS-1205190 would naturally lead to relatively weak [OIII] emission\footnote{While H$\beta$ EW does increase with decreasing metallicity, it only reaches $\sim$300 \AA{} at 0.01 $Z_{\odot}$ even in extremely rapidly star-forming systems (sSFR $\sim$ 100 Gyr$^{-1}$) according to the \citet{Gutkin2016} templates used in \textsc{beagle}.} even if it recently experienced a burst of star formation activity (sSFR $\gtrsim$ 30 Gyr$^{-1}$) as typically inferred for the strong \OIIIHb{} emitters in our sample.
If our above suspicions of COS-469100 and COS-1205190 are correct, all nine of our z$\simeq$7 \Lya{} detections are from sources with intense radiation fields and we would infer that such systems typically exhibit substantially stronger (5.9$^{\scaleto{+4.3}{4.5pt}}_{\scaleto{-3.1}{4.5pt}} \times$ higher EW) \Lya{} relative to the more typical massive z$\simeq$7 population which has moderate \OIIIHb{} (EW = 200--800 \AA{}), moderately-low metallicity ($\sim$0.2 $Z_{\odot}$; \citetalias{Endsley2021}), and emission dominated by star formation.

The strong \Lya{} emission (EW = 29 \AA{}) from our bluest source (COS-1205190) also motivates an investigation into how the observed \Lya{} EW is related to the rest-UV slope among massive z$\simeq$7 galaxies.
While it has been shown that \Lya{} becomes stronger in z$\simeq$4 galaxies from $\beta = -1.4$ to $\beta = -1.8$ \citep{Stark2010}, it is not necessarily clear that such a prominent trend would continue to exist when the bulk of the galaxy population is very blue ($\beta \lesssim -2$) as is the case at z$\simeq$7 \citep{Bouwens2014_beta}. 
To test for any such association, we split our sample (again ignoring COS-469110) into sources with $\beta < -2.1$ and $\beta > -2.1$, where we adopt a dividing point equal to the approximate median rest-UV slope of the entire sample ($\beta = -2.07$). 
Following the Bayesian approach used for the \OIIIHb{} vs. \Lya{} analysis above, we find that our bluest galaxies ($-3.4 < \beta < -2.1$) typically exhibit much stronger \Lya{} (4.0$^{\scaleto{+4.1}{4.5pt}}_{\scaleto{-1.9}{4.5pt}} \times$ higher EW) relative to those more representative of the massive z$\simeq$7 population ($-1.1 < \beta < -2.1$; \citetalias{Endsley2021}) as illustrated in the bottom panel of Fig. \ref{fig:z7EWDistnComparison}.
This trend likely reflects lower dust content (and therefore relatively little \Lya{} attenuation) within bluer sources as is the case at lower redshifts \citep[e.g.][]{Shapley2003,Pentericci2009,Kornei2010,Stark2010,Hathi2016,Trainor2016}. 
Notably, our z$\simeq$7 galaxies also follow a similar behavior to the z$\simeq$6 sources presented in \citet{deBarros2017} where our strongest \Lya{} emitters (EW$>$25 \AA{}) all possess very blue rest-UV slopes ($\beta < -2.1$; see Table \ref{tab:shortTable}) likely signaling particularly low dust content along the line of sight.

The overall picture emerging from these results is that the observed \Lya{} EW is substantially enhanced by a recent strong burst of star formation (or presence of an AGN) among massive reionization-era galaxies.
Low dust content along the line of sight will further enhance visible \Lya{} similar to trends at lower redshifts.
While we defer a more detailed physical interpretation until \S\ref{sec:discussion}, we emphasize that our results demonstrate that \Lya{} can be detected with high success rate in massive z$\simeq$7 galaxies possessing strong \OIIIHb{} emission ($>$800 \AA{} EW) and very blue rest-UV slopes ($\beta < -2.1$).

\begin{figure}
\includegraphics{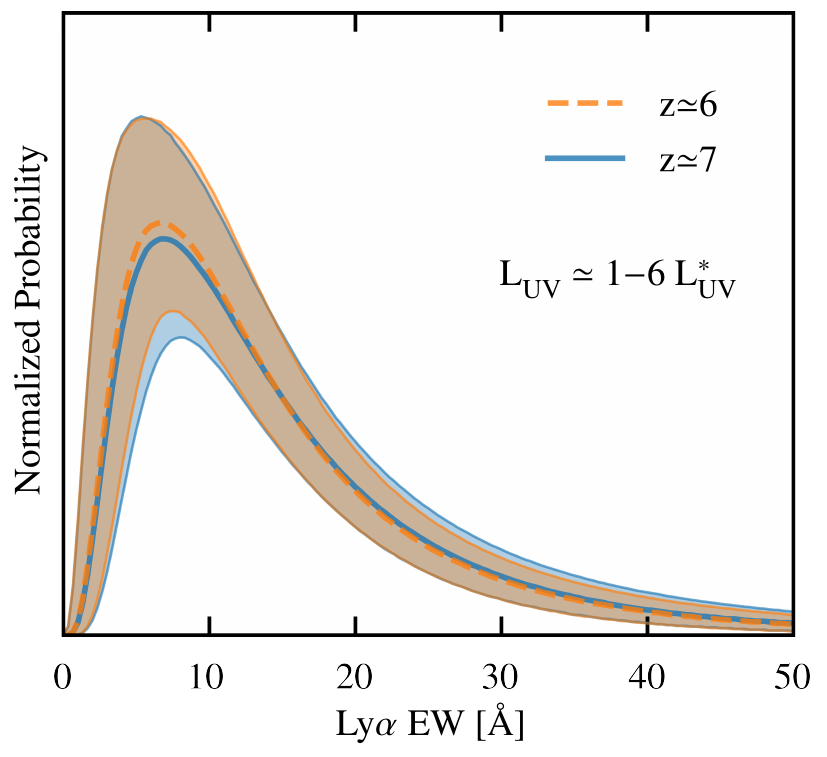}
\caption{Plot showing the inferred \Lya{} EW distributions of our massive galaxy samples at z$\simeq$7 (solid blue) and z$\simeq$6 (dashed orange) using the Bayesian approach of Eq. \ref{eq:prob2}. These two distributions are equivalent within uncertainties, suggesting that \Lya{} transmission is evolving less rapidly between z$\simeq$6--7 for the massive population relative to low-mass lensed systems \citep{Hoag2019,Fuller2020}. This is consistent with a scenario wherein massive z$\simeq$7 galaxies often reside in large highly-ionized bubbles.}
\label{fig:EW_evolution}
\end{figure}

\subsection{Evolution in the \Lya{} EW Distribution of Massive Galaxies from z$\simeq$6 to z$\simeq$7} \label{sec:evolution}

Over the past decade, a variety of observational campaigns have presented evidence that the IGM rapidly transitions from highly ionized at z$\simeq$6 (\xHI{} $\sim$ 10\%; e.g. \citealt{McGreer2015}) to substantially neutral at z$\simeq$7 (\xHI{} $\sim$ 50\%, e.g. \citealt{GreigMesinger2017}; \citealt{Zheng2017}; \citealt{Davies2018}; \citealt{Wang2020}; \citealt{Whitler2020}).
Such a rapid transition is naturally expected to reduce \Lya{} transmission between z$\simeq$6--7 for the typical galaxy population.
This is perhaps evidenced by the factor of $\sim$10 decline in the strong \Lya{} (EW$>$25 \AA{}) emitter fraction among the low-mass lensed population ($\sim$0.1 L$_{\mathrm{UV}}^{\ast}$; \citealt{Hoag2019,Fuller2020}).
One of the primary goals of our spectroscopic campaign is to build a sufficiently large sample of very bright z$\simeq$6--7 galaxies to explicitly test whether \Lya{} transmission from massive systems is evolving at a slower pace, as would be expected if they commonly reside in large ionized bubbles \citep[e.g.][]{Wyithe2005,McQuinn2007,Weinberger2018}.
While this campaign is still ongoing, we here report our current constraints on the \Lya{} EW distribution evolution between z$\simeq$6--7 using our sample of 30 and 22 massive (L$_{\mathrm{UV}}^{}$ $\simeq$ 1--6 L$_{\mathrm{UV}}^{\ast}$) galaxies at z$\simeq$6 and z$\simeq$7, respectively.

At each redshift, we infer the \Lya{} EW distribution following the Bayesian approach presented in \S\ref{sec:z7_LyA} where we assume a log-normal function parametrized by a median EW, \muEW{}, and standard deviation, \sigmaEW{}.
Here we also add a third parameter, $A$, quantifying the fraction of sources with \Lya{} EW$>$0 \AA{} because some galaxies may not show any \Lya{} in emission \citep[e.g.][]{Schenker2014,Mason2018_IGMneutralFrac}.
Using the same grid for \muEW{} and \sigmaEW{} as in \S\ref{sec:z7_LyA} and allowing $A$ to vary between 0--1 (with a spacing of 0.01), we calculate the probability for each set of parameters $P(\muEW{},\sigmaEW{},A)$ as in Eq. \ref{eq:prob}.

\begin{table}
\centering
\caption{Parameters inferred describing the log-normal \Lya{} EW distribution of massive galaxies at z$\simeq$7 and z$\simeq$6. The three parameters \muEW{}, \sigmaEW{}, and $A$ are, respectively, the median EW, standard deviation, and fraction of sources with \Lya{} EW$>$0 \AA{}.}
\begin{tabular}{P{1.0cm}P{1.5cm}P{1.5cm}P{1.5cm}} 
\hline
 & \muEW{} [\AA{}] & \sigmaEW{} [dex] & $A$ \Tstrut{} \Bstrut{} \\
\hline
z$\simeq$6 & 12.0$^{\scaleto{+2.8}{4.5pt}}_{\scaleto{-2.3}{4.5pt}}$ & 0.31$^{\scaleto{+0.11}{4.5pt}}_{\scaleto{-0.08}{4.5pt}}$  \Tstrut{}& 0.89$^{\scaleto{+0.08}{4.5pt}}_{\scaleto{-0.13}{4.5pt}}$ \\[6pt]
z$\simeq$7 & 11.0$^{\scaleto{+3.8}{4.5pt}}_{\scaleto{-3.2}{4.5pt}}$ & 0.37$^{\scaleto{+0.15}{4.5pt}}_{\scaleto{-0.10}{4.5pt}}$ & 0.88$^{\scaleto{+0.09}{4.5pt}}_{\scaleto{-0.15}{4.5pt}}$ \\[6pt]
\hline
\end{tabular}
\label{tab:EWDistn_params}
\end{table}

The \Lya{} EW distribution of our 22 massive z$\simeq$7 galaxies is well characterized by a log-normal distribution with a median EW \muEW{} = 11.0$^{\scaleto{+3.8}{4.5pt}}_{\scaleto{-3.2}{4.5pt}}$ \AA{} and standard deviation \sigmaEW{} = 0.37$^{\scaleto{+0.15}{4.5pt}}_{\scaleto{-0.10}{4.5pt}}$ dex.
The fraction of EW$>$0 \AA{} sources is found to be very high ($A$ = 0.88$^{\scaleto{+0.09}{4.5pt}}_{\scaleto{-0.15}{4.5pt}}$) with a best-fitting value of $A = 1$, suggesting that a large majority of massive z$\simeq$7 galaxies show \Lya{} in emission.
This EW distribution further implies that approximately 15\% of our z$\simeq$7 galaxies exhibit strong (EW$>$25 \AA{}) \Lya{}, in agreement with previous studies quantifying this \Lya{} emitter fraction among the bright (\Muv{} $<$ $-$20.25) population \citep[e.g.][]{Ono2012,Pentericci2018,Schenker2014}.
Interestingly, the \Lya{} EW distribution inferred from our 30 massive z$\simeq$6 galaxies is remarkably similar to that at z$\simeq$7 with all three parameters agreeing well within 1$\sigma$ uncertainties (see Table \ref{tab:EWDistn_params}).

Our results therefore appear to suggest that the \Lya{} EW distribution (and hence \Lya{} transmission) is evolving less rapidly for the massive population relative to low-mass galaxies \citep{Hoag2019,Fuller2020}.
To provide a more direct comparison, we now quantify the confidence to which our data rule out a strong (factor of $>$2) decline in \Lya{} transmission between z$\simeq$6--7.
We do so by assuming that massive galaxies have the same \Lya{} EW distribution at z$\simeq$6 and z$\simeq$7, with the exception that the EWs at z$\simeq$7 are multiplied by a factor $T$ which will describe evolution in \Lya{} transmission\footnote{This is similar to the `smooth' evolution approach of previous works \citep[e.g.][]{Treu2012,Pentericci2014}.}.
Adopting a log-normal EW distribution, the probability for a given set of parameters, $P(\muEW{},\sigmaEW{},A,T)$ is then calculated in a manner similar to Eq. \ref{eq:prob}:
\begin{equation} \label{eq:prob2}
\begin{split}
P (\muEW{},&\sigmaEW{},A,T) \propto \\
& \prod_i \int P_i\left(\mathrm{EW}\right) P\left(EW\ |\ \muEW{},\sigmaEW{},A\right)\  d\mathrm{EW} \, \, \, \, \times \\
& \prod_j \int P_j\left(\mathrm{EW}\right) P\left(EW\ |\ \muEW{},\sigmaEW{},A,T\right)\  d\mathrm{EW}.
\end{split}
\end{equation}
Here, $i$ and $j$ are the indices of the z$\simeq$6 and z$\simeq$7 sources, respectively.
We adopt the same grid of \muEW{}, \sigmaEW{}, and $A$ as above, as well as a grid in $T$ spanning 0--3 with a spacing of 0.01. 
Doing so, we infer $T$ = 1.04$^{\scaleto{+0.39}{4.5pt}}_{\scaleto{-0.29}{4.5pt}}$ which is consistent with unity as expected given the similarity of the EW distributions at z$\simeq$6 and z$\simeq$7 (Fig. \ref{fig:EW_evolution}).

Our goal is to quantify evolution in the transmission of \Lya{} for massive galaxies between z$\simeq$6--7.
While the value of $T$ inferred above does account for evolution in transmission, it also includes any evolution in the physical conditions that may be impacting \Lya{} production and escape within galaxies.
Fortunately, we can take the first steps towards decoupling these internal factors using the insight gained from our analysis in \S\ref{sec:z7_LyA}.
Therein, we found that the observed \Lya{} EW in massive z$\simeq$7 galaxies correlates with rest-UV slope and \OIIIHb{} EW. 
As detailed in \S\ref{sec:binospec}, the typical rest-UV slopes of our z$\simeq$6 ($\beta = -2.09$) and z$\simeq$7 ($\beta = -2.07$) samples are essentially equal.
Given this, we assume that the dust content of our two samples are similar enough to not cause a significant evolution in the \Lya{} EW distribution.
As discussed further in \S\ref{sec:discussion}, the trend between \OIIIHb{} EW and \Lya{} EW seen in our sample is likely driven by the relationship between \OIIIHb{} EW and the stellar ionizing photon production efficiency, \xiion{} \citep{Chevallard2018_z0,Tang2019}.
Our SED fits with \textsc{beagle} suggests that the typical ionizing photon production efficiencies of our z$\simeq$6 (\logxiion{} = 25.62) and z$\simeq$7 (\logxiion{} = 25.59) samples are also similar enough to not significantly impact the evolution of the \Lya{} EW distribution.
We therefore assume that our inferred value of $T$ is dominated by the evolution in \Lya{} transmission.

Equipped with this knowledge, we find that our spectroscopic data rule out a strong decline in \Lya{} transmission ($T_{\mathrm{z\simeq7}}$ / $T_{\mathrm{z\simeq6}} < 0.5$) with 98.5\% ($\approx$2$\sigma$) confidence. 
This is in contrast to the ten-fold decrease in strong \Lya{} emitters among low-mass lensed galaxies over the same redshift interval \citep{Hoag2019,Fuller2020}.
Therefore, while uncertainties are significant given our modest sample size, our results currently suggest that \Lya{} transmission is evolving less rapidly between z$\simeq$6--7 for the massive population.

There are a number of possible explanations for this difference in \Lya{} transmission evolution as we enter the epoch of reionization.
One is that massive z$\simeq$7 galaxies may tend to possess larger \Lya{} velocity offsets, helping push the photons into the damping wing before escaping the host galaxy \citep{Mason2018}.
This may be expected given the positive correlation between \Lya{} velocity offset and rest-UV luminosity at z$\sim$2--3 \citep{Erb2014}, likely reflecting (in part) the ability of more massive galaxies to drive stronger outflows.
Another explanation is that massive z$\simeq$7 galaxies are more likely to reside in large ionized bubbles \citep{Furlanetto2004,Wyithe2005,Lee2007,McQuinn2007,Weinberger2018} which enable \Lya{} photons to cosmologically redshift far into the damping wing and thus transmit more easily through a partially neutral IGM. 

\begin{figure}
\includegraphics{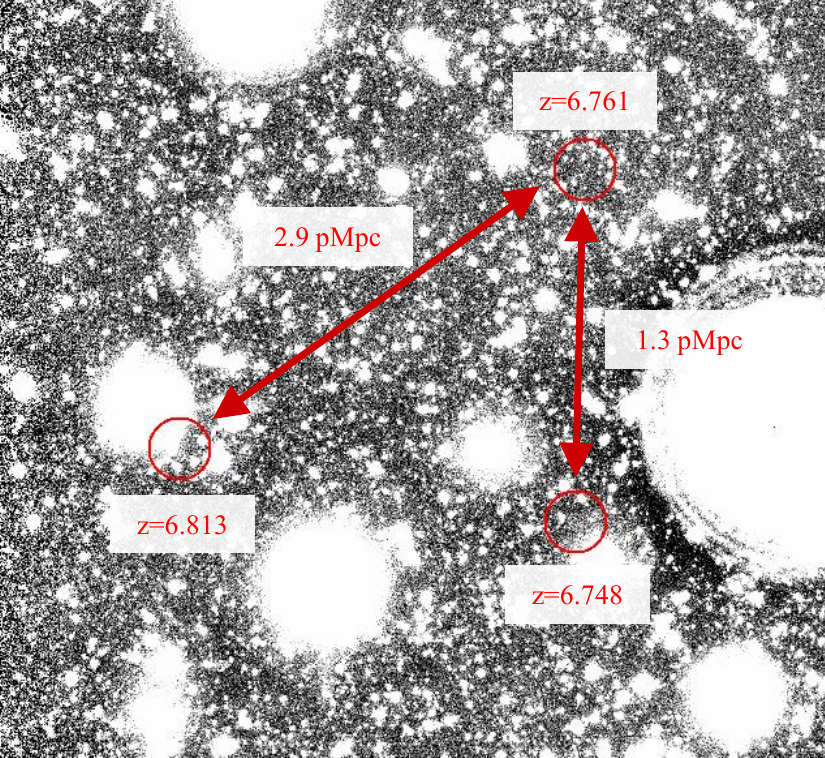}
\caption{Illustration of three closely separated \Lya{} emitters in our z$\simeq$7 sample. All three sources fall within a spherical region with radius $R = 1.7$ physical Mpc, consistent with the expected sizes of HII regions at z$\simeq$7 \citep[e.g.][]{Lin2016}. The proximity of these \Lya{} emitters thus may plausibly reflect the presence of a large ionized bubble.}
\label{fig:bubble}
\end{figure}

\subsection{A Possible Large Ionized Region at z$\simeq$7} \label{sec:bubbles}

As discussed in \S\ref{sec:evolution}, our results are consistent with a scenario wherein massive z$\simeq$7 systems often reside in large ionized regions.
In this same picture, we might expect to find some of our z$\simeq$7 \Lya{} emitters nearby one another if they reside in the same bubble. 
Interestingly, we do indeed find an instance where three of our massive z$\simeq$7 \Lya{} emitters possess very similar redshifts ($\Delta$z $\sim$ 0.05) and are also closely separated in angular space ($<$5 arcmin).

Our spectroscopic results indicate that COS-940214 (\zLya{} = 6.748), COS-1009842 (\zLya{} = 6.761), and COS-995126 (\zLya{} = 6.813) all lie within a spherical volume with radius $R = 1.7$ physical Mpc (Fig. \ref{fig:bubble}), consistent with the expected sizes of HII regions at z=7 \citep[e.g.][]{Furlanetto2004,Mesinger2007,Lin2016}.
These three sources have luminosities ranging from 0.8--2.3 L$^{\ast}_{\mathrm{UV}}$ and therefore are likely among the most massive galaxies present at z$\simeq$7.
All three are also inferred to be actively forming stars with SFRs of 7--28 \Msol{} yr$^{-1}$ using the SED fitting procedure described in \S\ref{sec:BEAGLE}.
Their very blue IRAC colours ([3.6]$-$[4.5] $<$ $-$0.6) further suggest large sSFRs ranging from approximately 10--130 Gyr$^{-1}$ (see Table \ref{tab:galaxy_properties}).
Because the typical sSFR at z$\simeq$7 is 5 Gyr$^{-1}$ (\citetalias{Endsley2021}), these three galaxies are likely undergoing a burst of star formation.

It is of interest to estimate whether these three z$\simeq$7 \Lya{} emitters could individually power ionized bubbles large enough to cover the separation between them.  
To estimate the plausible sizes of such bubbles, we assume that the dominant ionizing output from these three galaxies has been over the past 10 Myr as evidenced by their large sSFRs.
Because the recombination time-scale at z$\simeq$7 is much longer (of order the Hubble time\footnote{This is assuming a clumping factor of \textit{C}$\approx$3 at z=7  \citep[e.g.][]{Finlator2012,Shull2012,Pawlik2015}. More recent, higher resolution simulations find that the clumping factor is temporarily increased to \textit{C}$\sim$10--20 shortly ($\sim$3--10 Myr) after a region becomes reionized \citep{Park2016,DAloisio2020}. Adopting such a higher value would yield smaller estimated bubble sizes, still insufficient to cover the observed separation between the three galaxies.}), we can approximate the radius of an individual ionized region as \citep{Cen2000}:
\begin{equation}
    R \approx \left(\frac{3\ \dot{N}_{\mathrm{ion}}\ \fesc{}\ t}{4\pi\ \bar{n}_{_{\mathrm{HI}}}(z)}\right)^{1/3}
\end{equation}
Here, $t = 10$ Myr, $\bar{n}_{_{\mathrm{HI}}}(z)$ is the average hydrogen density at the redshift of each source, \fesc{} is the escape fraction of ionizing photons, and $\dot{N}_{\mathrm{ion}}$ is the rate at which ionizing photons are produced by stars in the galaxy.
We take $\dot{N}_{\mathrm{ion}}$ from the \textsc{beagle} SED fits to each galaxy and assume an escape fraction of \fesc{} = 20\% \citep[e.g.][]{Robertson2013}.
Our estimated bubble sizes are not significantly altered if we instead use the relation from \citet{Tang2019} to obtain \xiion{} from the inferred [OIII]$+$H$\beta$ EWs.

We estimate that the three aforementioned sources could individually power bubbles of sizes $R$ = 0.29--0.41 physical Mpc, consistent with other estimates recently reported at z$\simeq$7--8 \citep{Castellano2018,Tilvi2020}. 
The largest region ($R$ = 0.41 Mpc) is estimated to come from COS-955126 given that it is inferred to produce more than twice as many ionizing photons as the other two sources, mainly due to its much higher luminosity.
COS-940214 and COS-1009842 have very similar estimated bubble sizes ($R$ = 0.30 and 0.29 Mpc, respectively) due to the similarity in their inferred $\dot{N}_{\mathrm{ion}}$.
Notably, these estimated bubble sizes ($R$ $\sim$ 0.3 Mpc) are much smaller than the observed separation between the three galaxies ($R$ = 1.7 Mpc).

One way in which a larger bubble may have grown around these sources is through the impact of a local overdensity \citep[e.g.][]{Furlanetto2004,Wyithe2005,Lee2007,McQuinn2007}.
Such a scenario may already be hinted at by the proximity of these $\gtrsim$L$^{\ast}_{\mathrm{UV}}$ galaxies.
Using the z$\simeq$7 luminosity function from \citet{Bowler2017}, we would expect only N=0.7 bright ($\geq$0.8 L$^{\ast}_{\mathrm{UV}}$) galaxies on average within a 1.7 Mpc radius sphere. 
Because we know there are at least three such galaxies within this volume, our data suggest this region is likely a factor of $\gtrsim$4$\times$ overdense.
Further evidence of an overdensity comes from investigating the surface density of photometrically-selected z$\simeq$7 galaxies over this region.
In an $\approx$60 arcmin$^2$ rectangular area encompassing the three \Lya{} emitters, we identify a total of nine galaxies satisfying our z$\simeq$7 selection criteria (\S\ref{sec:sample_selection}), a factor of 6$\times$ above that expected on average\footnote{This $\approx$60 arcmin$^2$ rectangular area is entirely contained within an UltraVISTA ultra-deep stripe. We identify 67 z$\simeq$7 galaxies across all the ultra-deep stripes (0.73 deg$^2$ total) translating to an average surface density of 0.0255 arcmin$^{-2}$ (not corrected for completeness).} in the same area (N=1.5).
Future \Lya{} observations will enable us to better quantify the spectroscopic overdensity within this region.

\section{Discussion} \label{sec:discussion}

Numerous observational studies over the past decade have demonstrated a low ($\lesssim$10--20\%) \Lya{} detection rate among typical ($<$L$^{\ast}_{\mathrm{UV}}$) z$\gtrsim$7 galaxies \citep[e.g.][]{Ono2012,Treu2013,Pentericci2014,Pentericci2018,Schenker2014,Tilvi2014}.
It was therefore a surprise when \Lya{} was detected in all four of the brightest (3--4 L$^{\ast}_{\mathrm{UV}}$) z$>$7 galaxies selected over the CANDELS fields (i.e. the \citetalias{RobertsBorsani2016} sample), particularly given that the IGM is thought to be highly neutral at these epochs (\xHI{}$>$40\%; e.g. \citealt{Davies2018,Planck2020,Wang2020,Whitler2020}). 
The unusual \Lya{} detectability of these sources hence suggested that their \Lya{} photons may be less sensitive to strong scattering by the IGM, as would be expected if these objects reside in large, highly ionized bubbles \citep[e.g.][]{Zitrin2015,Weinberger2018}.
However, it has been proposed that intense radiation fields may also be driving their enhanced \Lya{} detectability given that all four \citetalias{RobertsBorsani2016} galaxies are inferred to possess strong \OIIIHb{} emission (EW $>$ 800 \AA{}; \citealt{Stark2017}).
In this section, we use our observations of a larger population (N=22) of similarly bright (L$^{}_{\mathrm{UV}}$ $\simeq$ 1--6 L$^{\ast}_{\mathrm{UV}}$) z$\simeq$7 galaxies with {\it Spitzer}/IRAC constraints on \OIIIHb{} EWs
to better understand the origin of strong \Lya{} within the \citetalias{RobertsBorsani2016} sample.

We find that  the detectability of \Lya{} depends strongly on the  \OIIIHb{} EW at z$\simeq$7.
We detect \Lya{} in 78\% (7/9) of bright z$\simeq$7 galaxies possessing strong \OIIIHb{} (EW $>$ 800 \AA{}) as opposed to only 8\% (1/12) of galaxies with more moderate \OIIIHb{} (EW = 200-800 \AA{}; \S\ref{sec:z7_LyA}). 
The much (4.7$\times$) higher \Lya{} EWs of the strong \OIIIHb{} population likely reflects, in part, a larger ionizing photon production efficiency associated with young stellar populations \citep{Chevallard2018_z0,Tang2019}. The high sSFRs ($\gtrsim$30 Gyr$^{-1}$) of these galaxies may also help drive low HI column density channels through the ISM/CGM \citep[e.g.][]{Clarke2002,Ma2020} through which \Lya{} photons can efficiently escape \citep{Jaskot2019,Gazagnes2020}. 

While we have shown that efficient ionizing photon production significantly boosted the \Lya{} detectability of the \citetalias{RobertsBorsani2016} sample, our results also suggest that the IGM ionization may have played a significant role in enhancing 
their visibility as well.
In particular, we find no evidence of strong evolution in the \Lya{} EW distribution of bright (L$^{}_{\mathrm{UV}}$ $\simeq$ 1--6 L$^{\ast}_{\mathrm{UV}}$) galaxies between z$\simeq$6--7 (\S\ref{sec:evolution}).
This result stands in contrast to the rapid downturn in the fraction of strong \Lya{} emitters ($>$25 \AA{} EW) among low-mass ($\sim$0.1 L$^{\ast}_{\mathrm{UV}}$) lensed galaxies between z$\simeq$6--7 \citep{Hoag2019,Fuller2020}, suggesting that Ly$\alpha$ in the bright population may evolve more slowly.
Such findings are consistent with theoretical expectation that the brightest (and hence the most massive; e.g. \citealt{BaroneNugent2014}) z$\simeq$7 galaxies trace overdensities that are the first to 
create large, highly-ionized bubbles \citep[e.g.][]{Furlanetto2004,Wyithe2005,Lee2007,McQuinn2007,Weinberger2018}.

If massive z$\simeq$7 galaxies do trace overdense regions, we may expect to find \Lya{} emitting galaxies in our sample nearby one another within the same ionized structure.
Indeed, our results revealed three z$\simeq$7 \Lya{} emitters separated by relatively small angular distances ($<$5 arcmin) and with similar redshifts (\S\ref{sec:bubbles}). 
The corresponding spatial separation of these \Lya{} emitters ($R$ = 1.7 physical Mpc) is consistent with the expected sizes of HII regions at z$\simeq$7 \citep{Furlanetto2004,Mesinger2007,Lin2016}, suggesting that they may lie within the same large, highly-ionized bubble.
None the less, with our present data, it is not yet possible to characterize the extent of the ionized regions surrounding these z$\simeq$7 \Lya{} emitters. 
There are two distinct possibilities consistent with our current data.
On one hand, these massive z$\simeq$7 galaxies may be tracing a large ionized region, with a size comparable to 
their spatial separation ($R \gtrsim$ 2 Mpc).  This would be expected if the 
galaxies trace a strongly-overdense structure with larger-than-average ionizing 
photon output.  In this case, we would expect to see enhanced  Ly$\alpha$ 
from all systems within the ionized region.  However it is also possible that these systems are situated within distinct, moderate-sized ($R \sim$ 0.3 Mpc) bubbles in which they are the dominant contributors of ionizing photons. 
The smaller bubbles would lead to larger damping wing attenuation of \Lya{} from the IGM.  But the corresponding reduction in the Ly$\alpha$ flux is  
countered by efficient production of nebular emission, as signaled by 
the large EW \OIIIHb{} emission (and hence high sSFRs). The transmission could 
be further boosted if these massive systems have 
large velocity offsets \citep[e.g.][]{Erb2014,Stark2017,Mason2018}, 
redshifting the line further into the damping wing before it 
encounters  hydrogen.

Which of these two pictures is true depends largely on whether 
the nearby Ly$\alpha$ emitting galaxies we have identified trace 
a strong overdensity in galaxies. 
In \S\ref{sec:bubbles}, we presented evidence that this may indeed be the case.
The number density of spectroscopically confirmed $>$0.8 L$^{\ast}_{\mathrm{UV}}$ systems in this region (N=3) is $\approx$4 
times the average (N=0.7) expected from z$\simeq$7 luminosity functions (e.g., \citealt{Bowler2017}).
Furthermore, the surface density of photometrically-selected z$\simeq$7 galaxies surrounding the nearby \Lya{} emitters is 6$\times$ the average.
This suggests that an overdense population of neighboring galaxies may potentially help power a large ionized region.

These results add to two similar instances of grouped \Lya{} emitting galaxies that have been reported at z=7.0--7.7 \citep{Vanzella2011,Castellano2018,Tilvi2020}.
Both these previously identified regions contain at least one bright ($>$L$^{\ast}_{\mathrm{UV}}$) galaxy.
Furthermore, one of these regions (the BDF) shows evidence of a host overdensity similar to that in our sample \citep{Castellano2016}.
Interestingly, deep spectroscopic follow-up of relatively faint ($<$0.7 L$^{\ast}_{\mathrm{UV}}$) z$\simeq$7 systems identified photometrically in the BDF revealed no \Lya{} detections \citep{Castellano2018}. 
As stated therein, this may suggest that many of the fainter systems reside outside the ionized region, though it is also possible that \Lya{} is preferentially seen in the brightest systems because of their  larger velocity offsets.
Ultimately, future \Lya{} spectroscopy of fainter sources (with known spatial positions via e.g. \JWST{} rest-optical line detections) will be able to discern whether the \Lya{} EW distribution in the vicinity of these massive \Lya{} emitters is enhanced towards larger values as would be expected if they reside in large bubbles.

\section{Summary} \label{sec:summary}
We present the first results from a new spectroscopic MMT/Binospec program aimed at targeting \Lya{} in a large sample of bright (L$^{}_{\mathrm{UV}}$ $\simeq$ 1--6 L$^{\ast}_{\mathrm{UV}}$) z$\simeq$7 galaxies selected over very wide-area fields ($\sim$3 deg$^2$ total).
We use these results to investigate to what extent strong \OIIIHb{} emission (and hence large sSFRs) boost \Lya{} detectability from bright reionization-era galaxies.  We secondly test whether the decline in the visibility of 
Ly$\alpha$ emission is less rapid for massive galaxies, as might be 
expected if they trace strong overdensities that are situated in 
large ionized bubbles. Our conclusions are as follows:

\begin{enumerate}
    \item From our 22 targeted bright z$\simeq$7 galaxies, we confidently ($>$7$\sigma$) detect \Lya{} from nine sources. The redshifts of our z$\simeq$7 \Lya{} emitters range from \zLya{}=6.650--7.093, consistent with expectations given our photometric selection criteria. We measure \Lya{} fluxes ranging from (5.1--18.6)$\times$10$^{-18}$ erg/s/cm$^2$ and rest-frame EWs spanning 3.7--43.1 \AA{}. For sources lacking a confident detection, the data typically place a (7$\sigma$) \Lya{} EW limit of $\lesssim$10 \AA{} in skyline-free regions of the spectra.
    
    \item We find that the detectability of \Lya{} depends strongly on \OIIIHb{} EW among luminous z$\simeq$7 galaxies. We confidently detect \Lya{} in 78\% (7/9) of sources with strong \OIIIHb{} emission (EW $>$800 \AA{}) as opposed to only 8\% (1/12) of sources with more moderate \OIIIHb{} (EW = 200-800 \AA{}). The much (4.7$\times$) higher \Lya{} EWs of the strong \OIIIHb{} population likely reflect a larger ionizing photon production efficiency (and hence likely also a larger L$_{\scaleto{\mathrm{Ly}\alpha}{6pt}}$/L$_{\scaleto{\mathrm{UV}}{4.5pt}}$) owing to extremely young, hot stars formed in a recent rapid upturn or burst in star formation activity \citep{Chevallard2018_z0,Tang2019}. The high sSFRs ($\gtrsim$30 Gyr$^{-1}$) of such galaxies may also help create low HI column density channels through the ISM/CGM \citep[e.g.][]{Clarke2002,Ma2020} through which \Lya{} photons can efficiently escape \citep[e.g.][]{Jaskot2019,Gazagnes2020}. Given the large variations in \Lya{} detectability within our data, it will be necessary to ensure that future samples be well matched in sSFR across cosmic time to well infer the evolution of the IGM neutral state.
    
    \item We tentatively detect the nebular NV$\lambda$1238.8,1242.8 doublet, a signpost of significant AGN activity, in one of our z$\simeq$7 \Lya{} emitters. This is the fifth tentative detection of NV so far reported in a z$\gtrsim$7 \Lya{} emitter \citep{Tilvi2016,Hu2017,Laporte2017,Mainali2018} suggesting that low-luminosity AGN are present in a subset of the most massive reionization-era galaxies. Further investigation of this population will ultimately help clarify the contribution of such AGN to cosmic reionization \citep{Madau2015,Giallongo2019,Grazian2020}.
    
    \item We find no evidence for strong evolution in the transmission of \Lya{} emission for the massive ($\simeq$1--6 L$^{\ast}_{\mathrm{UV}}$) population between z$\simeq$6--7 ($T_{\mathrm{z\simeq7}}/T_{\mathrm{z\simeq6}}$ = 1.04$^{\scaleto{+0.39}{4.5pt}}_{\scaleto{-0.29}{4.5pt}}$). 
    This is in contrast to observations of low-mass ($\sim$0.1 L$^{\ast}_{\mathrm{UV}}$) lensed galaxies which suggest a factor of $\sim$10 decline in transmission \citep{Hoag2019,Fuller2020}. With our current sample size, we can rule out a factor of $\geq$2 decline in \Lya{} transmission with 98.5\% ($\approx$2$\sigma$) confidence among massive galaxies. 
    We discuss a number of possible explanations for these findings, including the expectation that massive z$\simeq$7 galaxies often reside within large, highly-ionized bubbles \citep[e.g.][]{Furlanetto2004,McQuinn2007,Weinberger2018}.
    
    \item We find three \Lya{} emitters in our z$\simeq$7 sample separated by relatively small angular distances ($<$5 arcmin) and with nearly identical redshifts ($\Delta$z $\approx$ 0.05). These small angular separations may be a signpost of an ionized bubble enhancing \Lya{} transmission. Indeed, the spatial separations of these \Lya{} emitters ($R$ = 1.7 physical Mpc) are consistent with the expected sizes of HII regions at z$\simeq$7 \citep[e.g.][]{Furlanetto2004,Mesinger2007,Lin2016}. With our present data, we cannot yet characterize the full extent of the ionized regions surrounding these closely-separated \Lya{} emitters. However, we estimate that the these \Lya{} emitters are individually capable of powering ionized bubbles with radii of R$\sim$0.3 Mpc, consistent with other estimates recently reported at z$\simeq$7--8 \citep{Castellano2018,Tilvi2020}. Future 
    work targeting fainter galaxies in the region should be able to 
    determine if the ionized region extends beyond these radii.
    
    \item We find tentative evidence of an overdensity surrounding these closely-separated \Lya{} emitters. The number density of spectroscopically confirmed $>$0.8 L$^{\ast}_{\mathrm{UV}}$ systems in these regions (N=3) is $\simeq$4 times the average (N=0.7) expected from z$\simeq$7 luminosity functions. Furthermore, the surface density of photometrically-selected z$\simeq$7 galaxies surrounding the nearby \Lya{} emitters is 6$\times$ the average. Such an overdensity may help facilitate the growth of a large (R$\gtrsim$2 physical Mpc) bubble around these sources.
  
\end{enumerate}

As a next step towards understanding \Lya{} emission from massive reionziation-era galaxies, in future work we will combine this Binospec dataset with recent results from a Cycle 7 ALMA Large Program, the Reionization Era Bright Emission Line Survey (REBELS; 2019.1.01634.L). 
REBELS has begun providing systemic redshifts via the [CII]158$\mu$m emission line as well as constraints on far-infrared dust continuum emission for a number of our z$\simeq$7 targets.
With this more complete rest-UV through far-infrared view, we will begin to better characterize the \Lya{} velocity offsets of massive z$\simeq$7 galaxies as well as understand how their observed [CII] and dust continuum emission relates to \Lya{}. 

\section*{Acknowledgements}

RE and DPS acknowledge funding from JWST/NIRCam contract to the University of Arizona, NAS5-02015. JC and SC acknowledge financial support from the European Research Council (ERC) via an Advanced Grant under grant agreement no. 321323 -- NEOGAL. RJB and MS acknowledge support from the Nederlandse Organisatie voor Wetenschappelijk Onderzoek via TOP grant TOP1.16.057.
BER was supported in part by NASA program HST-GO-14747, contract NNG16PJ25C, and grant 80NSSC18K0563, and NSF award 1828315.
Observations reported here were obtained at the MMT Observatory, a joint facility of the University of Arizona and the Smithsonian Institution. 

This research has benefited from the SpeX Prism Library (and/or SpeX Prism Library Analysis Toolkit), maintained by Adam Burgasser at http://www.browndwarfs.org/spexprism. 
This research also made use of \textsc{astropy}, a community-developed core \textsc{python} package for Astronomy \citep{astropy:2013, astropy:2018}; \textsc{matplotlib} \citep{Hunter2007_matplotlib}; \textsc{numpy} \citep{van2011numpy}; and \textsc{scipy} \citep{jones_scipy_2001}.

\section*{Data Availability}
 
The optical through mid-infrared imaging data underlying this article are available through their respective data repositories. See \url{https://hsc-release.mtk.nao.ac.jp/doc/} for HSC data,  \url{http://www.eso.org/rm/publicAccess#/dataReleases} for UltraVISTA and VIDEO data, and \url{https://sha.ipac.caltech.edu/applications/Spitzer/SHA/} for IRAC data. The MMT/Binospec data will be shared upon reasonable request to the corresponding author.




\bibliographystyle{mnras}
\bibliography{paper_ref} 



\appendix

\section{IRAC Imaging} 

\begin{figure*}
\includegraphics{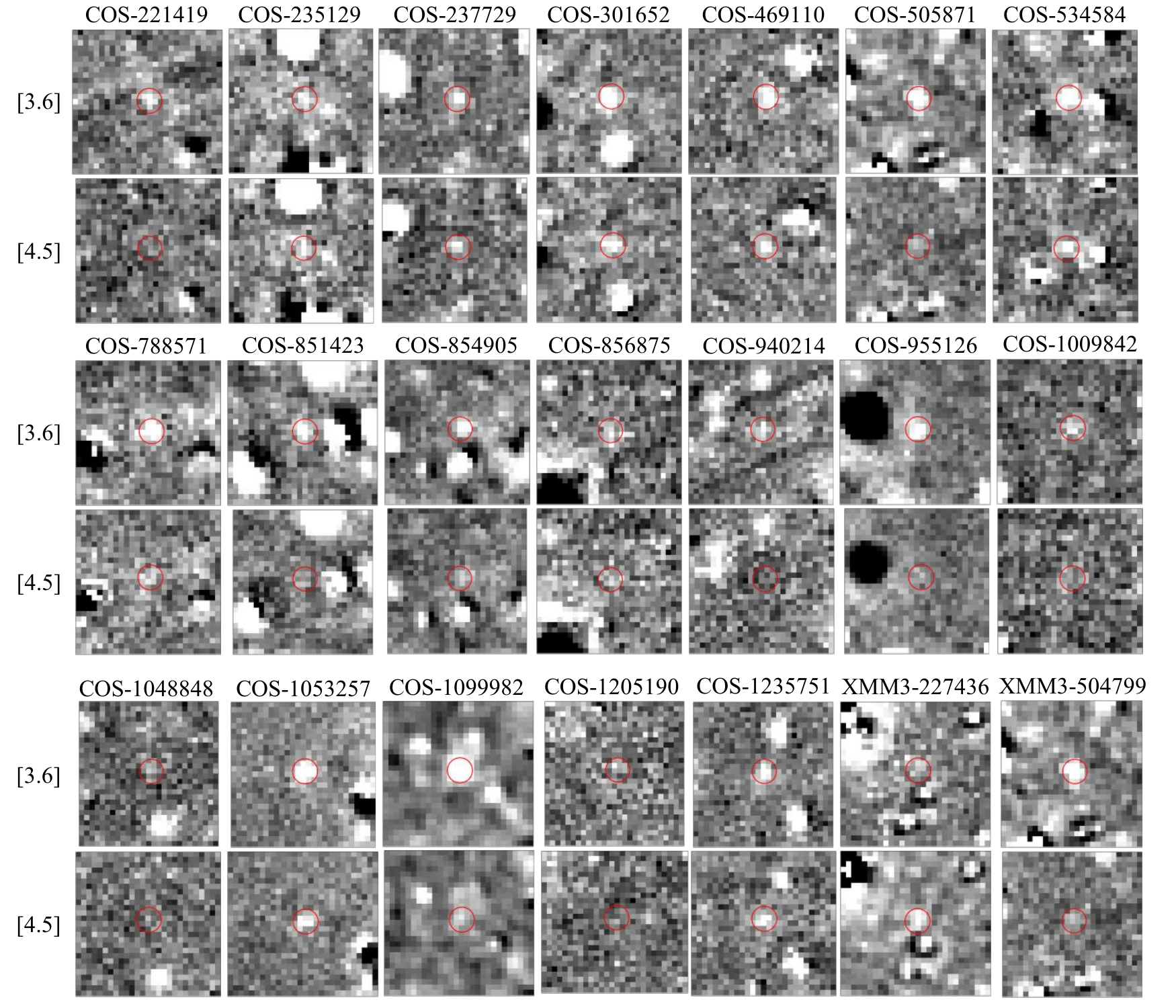}
\caption{Deconfused IRAC 3.6 and 4.5$\mu$m postage stamps of our z$\simeq$7 targets. Each postage stamp is approximately 16.5\arcsec{} by 16.5\arcsec{}. The red circle indicates the 2.8\arcsec{} diameter aperture used to measure the photometry. The postage stamps for COS-862541 are shown in \citetalias{Endsley2021}.}
\label{fig:IRAC}
\end{figure*}

We show the deconfused (\S\ref{sec:IRAC}) IRAC postage stamps of our z$\simeq$7 targets in Fig. \ref{fig:IRAC}. The postage stamps for COS-862541 are shown in \citetalias{Endsley2021}. 

\bsp	
\label{lastpage}
\end{document}